\documentclass[12pt,preprint]{aastex}
\usepackage{lscape}

\newcommand{\kms}{\ensuremath{\rm km\,s^{-1}}}
\newcommand{\ms}{\ensuremath{\rm m\,s^{-1}}}
\newcommand{\gcmc}{\ensuremath{\rm g\,cm^{-3}}}
\newcommand{\rhk}{\ensuremath{R^{\prime}_{\rm HK}}}	
\newcommand{\logrhk}{\ensuremath{\log\rhk}}		

\newcommand{\teff}{\ensuremath{T_{\rm eff}}}
\newcommand{\logg}{\ensuremath{\log{g}}}
\newcommand{\vsini}{\ensuremath{v \sin{i}}}
\newcommand{\feh}{[Fe/H]}

\newcommand{\rsun}{\ensuremath{R_\sun}}
\newcommand{\msun}{\ensuremath{M_\sun}}
\newcommand{\lsun}{\ensuremath{L_\sun}}

\newcommand{\rstar}{\ensuremath{R_\star}}
\newcommand{\mstar}{\ensuremath{M_\star}}

\newcommand{\lstar}{\ensuremath{L_\star}}

\newcommand{\rhostar}{\ensuremath{\rho_\star}}

\newcommand{\teq}{\ensuremath{T_{\rm eq}}}

\newcommand{\koi}{K00070}
\newcommand{\koic}{K00070.01}
\newcommand{\koib}{K00070.02}
\newcommand{\koid}{K00070.03}
\newcommand{\koie}{K00070.04}
\newcommand{\koif}{K00070.05}
\newcommand{\starname}{Kepler-20}
\newcommand{\planetb}{Kepler-20b}
\newcommand{\planetc}{Kepler-20c}
\newcommand{\planetd}{Kepler-20d}

\newcommand{\kicid}{KIC~6850504}
\newcommand{\tmid}{2MASSJ19104752+4220194}

\newcommand{\kicra}{\ensuremath{19^{\mathrm{h}}10^{\mathrm{m}}47^{\mathrm{s}}.52}}
\newcommand{\kicdec}{\ensuremath{+42^{\circ}20'19''.4}}
\newcommand{\kepmag}{12.498}

\newcommand{\rmag}{12.423}

\newcommand{\kepler}{\emph{Kepler}}
\newcommand{\spitzer}{\emph{Spitzer}}
\newcommand{\wspitzer}{Warm \emph{Spitzer}}
\newcommand{\blender}{{\tt BLENDER}}

\shortauthors{T.~N.~Gautier et al.}
\shorttitle{The Kepler-20 Planetary System}

\begin{document}

\bibliographystyle{apj}

\title{Kepler-20: A Sun-like Star with Three Sub-Neptune Exoplanets and Two Earth-size Candidates}

\author{
Thomas~N.~Gautier III\altaffilmark{1},
David Charbonneau\altaffilmark{2},
Jason F. Rowe\altaffilmark{3},
Geoffrey W. Marcy\altaffilmark{4},
Howard Isaacson\altaffilmark{4},
Guillermo Torres\altaffilmark{2}, 
Francois Fressin\altaffilmark{2},
Leslie A. Rogers\altaffilmark{5},
Jean-Michel D{\'e}sert\altaffilmark{2},
Lars A. Buchhave\altaffilmark{6,7},
David W. Latham\altaffilmark{2},
Samuel N. Quinn\altaffilmark{2},
David R. Ciardi\altaffilmark{8},
Daniel C. Fabrycky\altaffilmark{9},
Eric B. Ford\altaffilmark{10},
Ronald L. Gilliland\altaffilmark{11},
Lucianne M. Walkowicz\altaffilmark{12},
Stephen T. Bryson\altaffilmark{3},
William D. Cochran\altaffilmark{13},
Michael Endl\altaffilmark{13},
Debra A. Fischer\altaffilmark{14},
Steve B. Howell\altaffilmark{3},
Elliott P. Horch\altaffilmark{15},
Thomas Barclay\altaffilmark{16}
Natalie Batalha\altaffilmark{17},
William J. Borucki\altaffilmark{3},
Jessie L. Christiansen\altaffilmark{3},
John C. Geary\altaffilmark{2},
Christopher E. Henze\altaffilmark{3},
Matthew J. Holman\altaffilmark{2},
Khadeejah Ibrahim\altaffilmark{3},
Jon M. Jenkins\altaffilmark{18},
Karen Kinemuchi\altaffilmark{16},
David G. Koch\altaffilmark{3},
Jack J. Lissauer\altaffilmark{3},
Dwight T. Sanderfer\altaffilmark{3},
Dimitar D. Sasselov\altaffilmark{2},
Sara Seager\altaffilmark{5},
Kathryn Silverio\altaffilmark{4},
Jeffrey C. Smith\altaffilmark{18},
Martin Still\altaffilmark{16},
Martin C. Stumpe\altaffilmark{18},
Peter Tenenbaum\altaffilmark{18},
Jeffrey Van Cleve\altaffilmark{18}
}

\altaffiltext{1}{Jet Propulsion Laboratory/California Institute of Technology, Pasadena, CA 91109; thomas.n.gautier@jpl.nasa.gov}
\altaffiltext{2}{Harvard-Smithsonian Center for Astrophysics, Cambridge, MA 02138}
\altaffiltext{3}{NASA Ames Research Center, Moffett Field, CA 94035}
\altaffiltext{4}{Department of Astronomy, University of California, Berkeley, CA 94720}
\altaffiltext{5}{Massachusetts Institute of Technology, Cambridge, MA 02139}
\altaffiltext{6}{Niels Bohr Institute, University of Copenhagen, DK-2100, Copenhagen, Denmark}
\altaffiltext{7}{Centre for Star and Planet Formation, Natural History Museum of Denmark, University of Copenhagen, DK-1350, Copenhagen, Denmark}
\altaffiltext{8}{NASA Exoplanet Science Institute/California Institute of Technology, Pasadena, CA 91125}
\altaffiltext{9}{Department of Astronomy and Astrophysics, University of California, Santa Cruz, CA 95064}
\altaffiltext{10}{Astronomy Department, University of Florida, Gainesville, FL 32111}
\altaffiltext{11}{Department of Astronomy, 525 Davey Lab, The Pennsylvania State University, University Park, PA  16802}
\altaffiltext{12}{Department of Astrophysical Sciences, Princeton University, Princeton, NJ 08544}
\altaffiltext{13}{McDonald Observatory, The University of Texas at Austin, Austin, TX 78712}
\altaffiltext{14}{Department of Astronomy, Yale University, New Haven, CT 06511}
\altaffiltext{15}{Department of Physics, Southern Connecticut State University, New Haven, CT 06515}
\altaffiltext{16}{Bay Area Environmental Research Institute/NASA Ames Research Center, Moffett Field, CA 94035}
\altaffiltext{17}{Department of Physics and Astronomy, San Jose State University, San Jose, CA 95192}
\altaffiltext{18}{SETI Institute/NASA Ames Research Center, Moffett Field, CA 94035}

\begin{abstract}
We present the discovery of the \starname\ planetary system, which we initially identified through the detection of
five distinct periodic transit signals in the \kepler\ light curve of the host star \tmid. From high-resolution spectroscopy of the star, we
find a stellar effective temperature $\teff=5455\pm100$~K, a metallicity of \feh$=0.01\pm0.04$, and a surface gravity of $\logg=4.4\pm0.1$.
We combine these estimates with an estimate of the stellar density
derived from the transit light curves to deduce a stellar mass
of $\mstar=0.912\pm0.034~\msun$ and a stellar radius of $\rstar=0.944^{+0.060}_{-0.095}~\rsun$.  For three of the transit signals, 
we demonstrate that our results strongly disfavor the possibility
that these result from astrophysical false positives.  We accomplish this by first identifying the subset of stellar blends that reproduce 
the precise shape of the light curve and then using the constraints on the presence of additional stars from 
high-angular resolution imaging, photometric colors, and the absence of a secondary
component in our spectroscopic observations. We conclude that the planetary scenario is more likely than that of an astrophysical false positive 
by a factor of $2\times10^5$ (\planetb), $1\times10^5$ (\planetc), and $1.1\times10^3$ (\planetd), sufficient to validate these objects as planetary companions.  For \planetc\ and \planetd, the blend scenario is independently disfavored by the achromaticity of the transit: From \spitzer\ data gathered at
4.5~\micron, we infer a ratio of the planetary to stellar radii of $0.075\pm0.015$ (\planetc) and $0.065\pm0.011$ (\planetd), 
consistent with each of the depths measured in the \kepler\ optical bandpass.
We determine the orbital periods and physical radii of the three confirmed planets to be $3.70$~d and $1.91^{+0.12}_{-0.21}~R_{\Earth}$ 
for \planetb, $10.85$~d and $3.07^{+0.20}_{-0.31}~R_{\Earth}$ for \planetc, and $77.61$~d and $2.75^{+0.17}_{-0.30}~R_{\Earth}$ 
for \planetd.  From multi-epoch radial velocities, we determine the masses of \planetb\ and \planetc\ to
be $8.7\pm2.2~M_{\Earth}$ and $16.1\pm3.5~M_{\Earth}$, respectively, and we place an upper limit on the mass of \planetd\ 
of $20.1~M_{\Earth}~(2~\sigma)$.  
\end{abstract}



\section{Introduction}
    Systems with multiple exoplanets, and transiting exoplanets, each bolster confidence in the reality 
of the planetary interpretation of the signals and offer distinct constraints on models of planet formation.

    The first extrasolar planets were found around a pulsar \citep{wols92}, and it was the multi-planetary nature --- 
in particular the gravitational perturbations between the planets \citep{rasio92, wols94} --- 
which solidified this outlandish claim.  Around Sun-like stars as well, the origin of radial velocity signals continued to 
be questioned by some, at the time multiple planets were found around ups Andromeda \citep{butler99}.  
The orbital configuration of planets relative to each other has shed light on a host of physical processes, from 
smooth radial migration into resonant orbits \citep{lee02} to chaotic scattering into secular eccentricity cycles 
\citep{malh02,ford05}.  Now with ever-growing statistics of ever-smaller Doppler-detected planets in multiple 
systems \citep{mayor11}, the formation and early history of planetary systems continues to come into sharper focus.

    Concurrently, transiting exoplanets have paid burgeoning dividends, starting with the definitive proof that Doppler 
signals were truly due to gas-giant planets orbiting in close-in orbits \citep{charbonneau00, henry00}.  
Transit lightcurves offer precise geometrical constraints on the orbit of the planet \citep{winn10}, such that radial 
velocity and photometric measurements yield the density of the planet and hence point to its composition 
\citep{adams08, miller11}.   Transiting configurations also enable follow-up measurements 
\citep{charbonneau02, knutson07, triaud10} which inform on the mechanisms of planetary formation, evolution, 
and even weather.

    These two research streams, multiplanets and transiting planets, came together for the first time with the discovery of 
Kepler-9 \citep{holman10, torres11}. This discovery was enabled by data from the \kepler\ Mission
\citep{borucki10, koch10a}, which is uniquely suited for such detections as it offers near-continuous high-precision photometric monitoring 
of target stars.  Based on the first 4 months of \kepler\ data, \citet{borucki11} announced the detection of 170 stars each with 
2 or more candidate transiting planets; \citet{steffen10} discussed in detail 5 systems each possessing multiple candidate transiting planets.  
A comparative analysis of the population of candidates with multiple planets and single planets
was published by \citet{latham11}, and \citet{lissauer11a} discussed
the architecture and dynamics of the ensemble of candidate multi-planet systems.

The path to confirming the planetary nature of such \kepler\ candidates is
arduous.  At present, three stars (in addition to Kepler-9) hosting multiple transiting candidates have been presented in detail and the planetary
nature of each of the candidates has been established: These systems are Kepler-10 \citep{batalha11, fressin11a}, 
Kepler-11 \citep{lissauer11b}, and Kepler-18 \citep{cochran11}.  Transiting planets are most profitable when their masses
can be determined directly from observation, either through radial velocity (RV) monitoring of the host star or by transit timing variations (TTVs),
as was done for Kepler-9bc, Kepler-10b, Kepler-11bcdef, and Kepler-18bcd.  
When neither the RV or TTV signals is detected, statistical arguments
can be employed to show that the planetary hypothesis is far more likely than alternate scenarios (namely blends of several stars 
containing an eclipsing component), and this was
the means by which Kepler-9d, Kepler-10c, and Kepler-11g were all validated.  While such work proves the existence of a planet and determines
its radius, the mass and hence composition remain unknown save for speculation from theoretical considerations.

This paper presents the discovery of a new system, \starname, with five candidate transiting planets. 
We validate three of these by statistical argument; 
we then proceed to use RV measurements to determine the masses of two of these, and we
place an upper limit on the mass of the third. We do not validate in
this paper the remaining two signals (and hence remain only
candidates, albeit very interesting ones, owing to their diminutive
sizes), rather the validation of these two remaining signals is addressed in a separate
effort \citep{fressin12}.  The paper is structured as follows: In \S2, we present our extraction
of the \kepler\ light curve (\S2.1), our modeling of these data (and RVs) to estimate the orbital and physical parameters of the planets 
and star (\S2.2), as well as limits on the motion of the photocentroid during transit (\S2.3) and a study of the long-term astrophysical
variability of the star from the \kepler\ light curve (\S2.4).  In \S3 we present
follow-up observations that we use to argue for the planetary interpretation, including high-resolution imaging (\S3.1) and 
\spitzer\ photometry (\S3.2), and the spectroscopy we use to characterize the star and determine the radial velocity signal (\S3.3). 
In \S4, we present our statistical analysis that validates the planetary nature of the three largest candidate planets in the system. In \S5 we consider the dynamics of the system, and in \S6 we discuss the constraints on the composition and formation history of the three planets.

\subsection{Nomenclature}
Throughout the course of the \kepler\ Mission, a given star is known by many different names \citep[see][for an explanation of \kepler\ naming conventions]{borucki11}, and we pause here to explain the relationship of these names in the current context.  The star that is the subject of this paper is located at $\alpha=$\kicra, $\delta=$\kicdec\ (J2000).  It was already known as \tmid, and in the \kepler\ input catalog it was designated \kicid.  After the identification of candidate transiting planets it became a \kepler\ Object of Interest (KOI) and was further dubbed \koi, and it appeared as such in the list of candidates published by \citet{borucki11}. Some authors have elected to denote KOIs using a different nomenclature, in which case \koi\ would be identified as KOI-70. After the confirmation of the planetary nature of three of these candidates it was given its final moniker \starname. This paper describes that process of confirmation, but for simplicity we refer to the star as \starname\ throughout.  The three confirmed exoplanets were initially assigned KOI designations representing the chronological order in which the transiting signals were identified, but to avoid confusion we will refer to them henceforth by their \starname\ designations in which they are ordered by increasing orbital period $P$; \planetb\ (\koib, $P=3.70$~d), \planetc\ (\koic, $P=10.85$~d), and \planetd\ (\koid, $P=77.61$~d).  We will refer to the two remaining candidates as \koie\ and \koif, but note (as described below) that the period of \koie\ ($P=6.10$~d) is intermediate between those of \planetb\ and \planetc, and the period of \koif\ ($P=19.58$~d) is intermediate between those of \planetc\ and \planetd.

\section{Kepler Photometry and Analysis}

\subsection{Light Curve Extraction}

\begin{figure}
\includegraphics[scale=0.65]{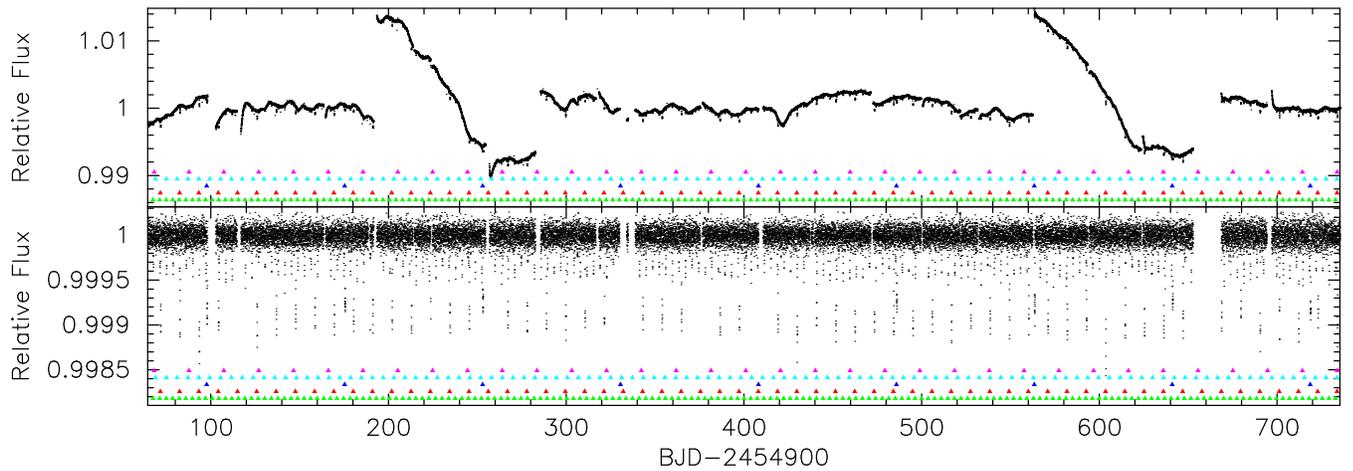}\\
\caption{\kepler\ light curve of \starname\ at a cadence of
  30~minutes. {\it Upper panel:} The normalized raw SAP light curves
  for Q1$-$Q8.  The star is positioned on one of four
  different detectors, depending upon the particular quarter, which
  results in the most obvious offsets that occur roughly 4 times per
  year. The other discontinuities are due to effects such as
  spacecraft safe-mode events and loss of fine pointing. 
   {\it Lower panel:} The SAP light curve after removing instrumental and long-term 
   astrophysical variability via polynomial fitting (see \S\ref{sec:model}).
   In both panels transits of Kepler-20b are marked in green, 20c in red, 20d in blue, 
   K00070.04 in cyan and K00070.05 in magenta.}
\label{fig:lc_unphased}
\end{figure}

\kepler\ observations of \starname\ commenced UT 2009 May 13 with Quarter 1 (Q1), and the \kepler\ data that we describe here extend through UT 2011 March 14 corresponding to the end of Quarter 8 (Q8), resulting in near-continuous monitoring over a span of 22.4~months. The \kepler\ bandpass spans $423$ to $897$~nm for which the response is greater than 5\% \citep{vancleve09}. This wavelength domain is roughly equivalent to the $V+R$-band \citep{koch10a}.  These observations have been reduced and calibrated by the \kepler\ pipeline \citep{jenkins10a}.   The \kepler\ pipeline produces calibrated light curves referred to as Simple Aperture Photometry (SAP) data in the 
\kepler\ archive, and this is the data product we used as the initial input for our analysis to determine the system parameters (see below). The pipeline provides time series with times in Barycentric Julian Days (BJD), and flux in photo-electrons per observation. The data were initially gathered at long cadence \citep{caldwell10,jenkins10b} consisting of an integration time per data point of 29.426~minutes.  After the identification of candidate transiting planets in the data from Q1, the target was also observed at short cadence \citep{gilliland10} corresponding to an integration time of 1~minute for Q2$-$Q6. We elected to use the long cadence version of the entire Q1$-$Q8 time series for computational efficiency.  There are 29,595 measurements in the Q1$-$Q8 time series. The upper panel of Figure~\ref{fig:lc_unphased} shows the raw \kepler\ Q1$-$Q8 light curve of \starname.  The data are available electronically from the Multi Mission Archive at the Space Telescope Science Institute (MAST) Web site\footnote{\tt http://archive.stsci.edu/kepler}.

\subsection{Derivation of System Parameters}\label{sec:model}

The five candidate transiting planets that are the subject of the paper were identified by the procedure described in \citet{borucki11}. 
Four of them (\planetb, \planetc, \planetd, \koie) are listed in that paper, and \koif\ was detected subsequently. 

We first cleaned the 
Q1$-$Q8 long-cadence \kepler\ SAP photometry of \starname\ of instrumental and long-term astrophysical variability not 
related to the planetary transits by fitting and removing a second-order polynomial to each contiguous photometric segment. We 
defined a segment to be a series of long-cadence observations that does not have a gap larger than five measurements (spanning at least 
2.5 hours).  In this process, we gave no statistical weight to observations that fell within a transit. We then normalized the 
corrected light curve by its median, and this cleaned light curve is displayed in the lower panel of Figure~\ref{fig:lc_unphased}. 

We then modeled simultaneously both this cleaned photometric time series and the radial-velocity measurements (described below 
in \S\ref{sec:spectroscopy}, and listed in Table~\ref{tab:rvtab}) to estimate the orbital and physical parameters of the star and its 
candidate planets. The free parameters in the fit were the mean stellar density $\rho_{\star}$ and the radial velocity instrumental zero point 
$\gamma$, and 7 parameters for each of the 5 planet candidates $i=\{$\planetb, \planetc, \planetd, \koie, \koif$\}$, namely the epoch of 
center of transit $T_{0,i}$, the orbital period $P_i$, the impact parameter $b_i$, the ratio of the planetary and stellar radii $(R_{\rm p}/R_{\star})_i$, 
 the radial velocity semi-amplitude $K_i$, and 
the two quantities, $(e\ \cos{\omega})_i$ and $(e\ \sin{\omega})_i$, relating the eccentricity $e_i$ and the argument of pericenter 
$\omega_i$.  The ratios of the semi-major axes to the stellar radius, $(a/{R_\star})_i$, were calculated from $\rho_{\star}$ and the orbital periods $P_i$ 
assuming $e = 0$ and that $M_{\star} \gg$ sum of the planet masses. 
(We note that our observations do not constrain the eccentricity, but we include it so that our error estimates of the 
other parameters are inflated to account for this possibility. Similarly, we are not able to detect the radial-velocity signals $K_i$ 
for \planetd, \koie, or \koif, but by including these parameters, we include any inflation these may imply for the uncertainties on the 
mass estimates of \planetb\ and \planetc, and the upper limit on the mass of \planetd.) 

We computed each transit shape using the analytic 
formulae of \citet{mandel02}.  We adopted a fourth-order non-linear limb-darkening law with coefficients fixed to those presented by 
\citet{claret11} for the \kepler\ bandpass using the parameters \teff, \logg, and \feh\ determined from spectroscopy 
(\S\ref{sec:spectroscopy}). Our approach implicitly assumes that all 5 transit signals are due to planets orbiting \starname; the validation 
of the 3 largest planets, \planetb, \planetc, and \planetd, is presented in \S4.  Using the validation approach presented in \S4, we 
are not able to validate the remaining two candidates \koie\ and \koif. Instead this difficult problem is deferred to a subsequent 
study \citep{fressin12}. We further assumed that the planets followed non-interacting Keplerian orbits, and that the eccentricity of each 
planetary orbit was constrained to be less than the value at which it would cross the orbit of another planet, $e \le 0.396$ (\planetb), 
$0.319$ (\planetc), $0.601$ (\planetd), $0.283$ (\koie), and $0.325$ (\koif).  Finally, we included an additional error term on the radial 
velocities (beyond those appearing in Table~\ref{tab:rvtab}) with a typical amplitude of 2~\ms, to assure that we were not underestimating 
the uncertainties on the radial velocities (and hence the planetary masses).

We included a prior on $\rho_{\star}$ as follows.  We matched the 
spectroscopically-estimated \teff, \logg, and \feh\ and the corresponding error estimates (see \S\ref{sec:spectroscopy}) to 
the Yonsei-Yale evolution tracks \citep{yi01, demarque04} with a Markov-Chain Monte Carlo (MCMC) routine to produce 
posterior distributions of stellar mass \mstar\ and stellar radius \rstar\ which in turn we used to generate the prior on \rhostar\ used 
for the determination of the best-fit model. The posterior distributions of $(a/{R_\star})_i$ were obtained by calculating $(a/{R_\star})_i$ 
for each element of the Markov chain.

We identified the best-fit model by minimizing the ${\chi}^2$ 
statistic using a Levenberg-Marquardt algorithm.  We estimated the uncertainties via the construction of a co-variance matrix (these 
results were also used below in the estimate of the width of the Gibbs sample for our MCMC analysis).  We then adopted the best-fit model 
(and its estimated uncertainties) as the seed for an MCMC analysis to determine the posterior distributions of all the model parameters.  We 
used a Gibbs sampler to identify new jump values of test parameter chains by drawing from a normal distribution.  The width of the normal 
distribution for each fitted parameter was initially determined by the error estimates from the best-fit model.  We generated 500 elements in 
the chain and then stopped to examine the success rate, and then were scaled the normal distributions using Equation~8 from 
\citet{gregory11}.  We repeated this process until the success rate for each parameter fell between 22$-$28\%.  We then held the width of 
the normal distribution fixed. 

To handle the large correlation between the model parameters, 
we adopted a hybrid MCMC algorithm based on Section 3 of \citet{gregory11}. The routine works by randomly using a Gibbs sampler 
or a buffer of previously computed chain points to generate proposals for a jump to a new location in the parameter space in a manner similar 
to the DE-MC algorithm \citep{terbrakk06}. The addition of the buffer allows for a calculation of vectorized jumps, which permits efficient 
sampling of highly correlated parameter space.  Once the proposals drawn from the buffer reached an acceptance rate that fell between 
$22-28$\%, we held the buffer fixed.  With the widths of the Gibbs sampler and buffer contents stabilized, we generated 4 chains, 
each with 1,000,000 elements. The generation of 4 separate chains permitted us to test for convergence via a Gelman-Rubin test. 

\begin{figure}
\begin{center}
\includegraphics[scale=1.0]{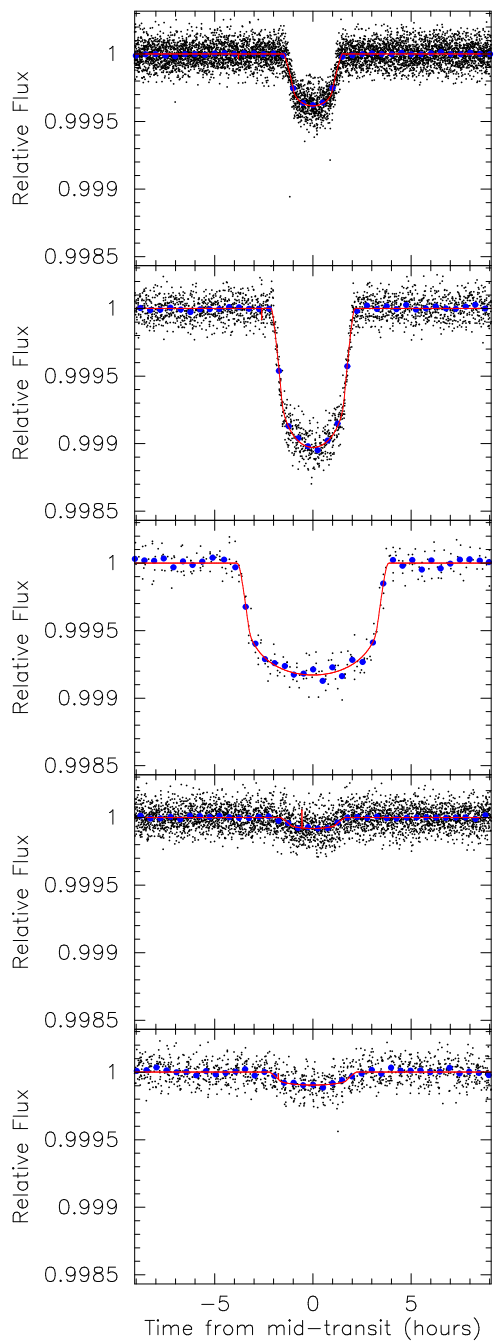}
\caption{\kepler\ light curves with an observational cadence of 30~minutes {\it (black points)} of \starname, phased to each of the periods of the 5 
candidate transiting planets (only data in the vicinity of each phased transit are shown).  Kepler-20b, 20c, 20d, K00070.04 and K00070.05 are shown 
from top to bottom. Blue points with error bars show these measurements binned in phase in increments of 30~minutes.  The red curve shows the 
global best-fit model (see \S2.2), which includes smoothing to match this 30~minute cadence. }
\label{fig:lc_phased}
\end{center}
\end{figure}

\begin{figure}
\begin{center}
\includegraphics[angle=270,scale=0.7]{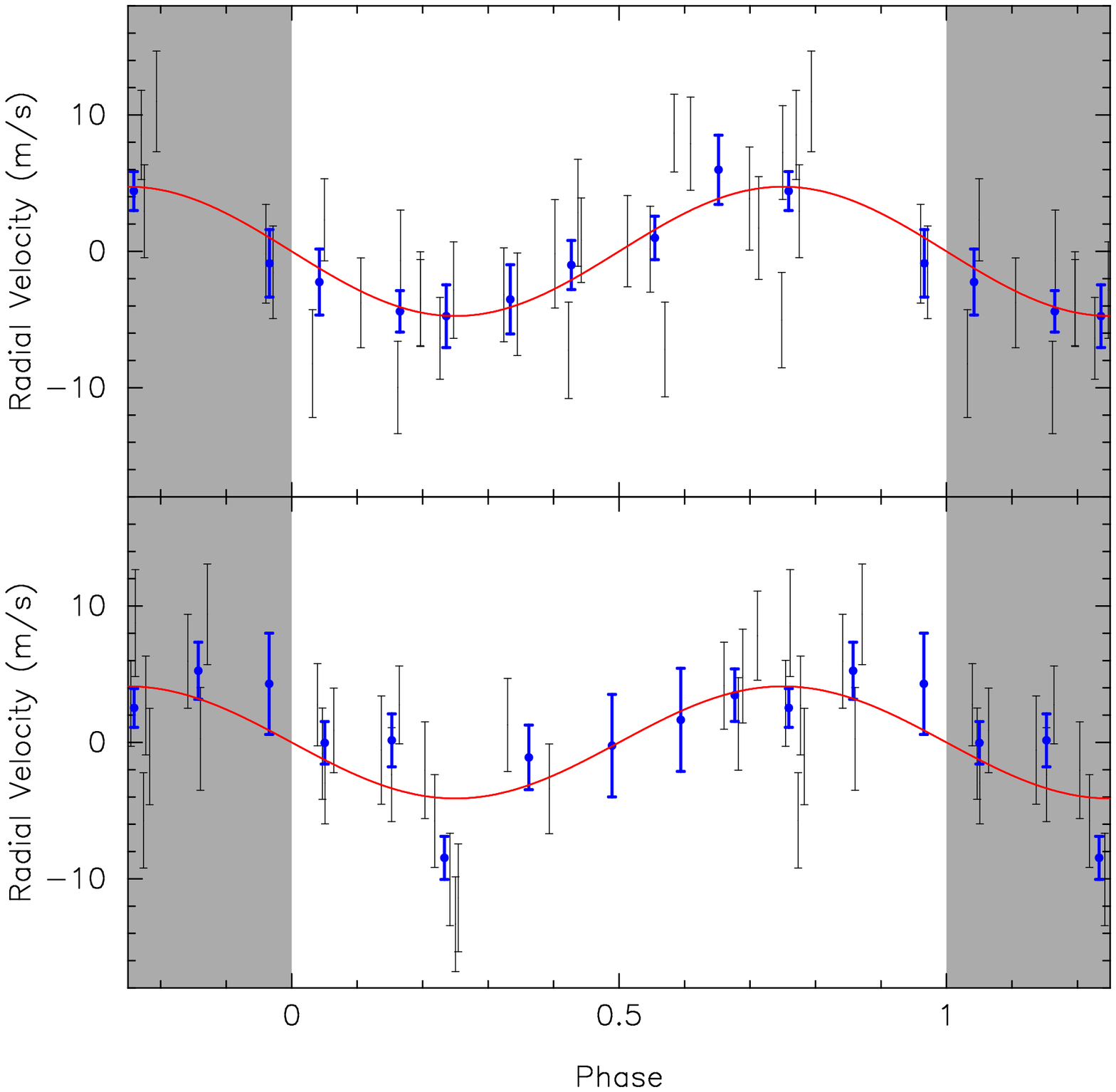}
\caption{{\it Upper Panel:} Radial velocities of \starname\ after correcting for the best-fit amplitudes of \planetb, \planetd, \koie, and \koif, 
leaving the effect of only \planetc\ and plotted as a function of its
orbital phase of \planetc.  Individual measurements as shown as gray points and these values binned in increments of 0.1 phase units
are shown in blue.  The phase coverage is extended by 0.25 phase units
on either side to show data continuity, but it should be noted the values in these gray regions are plotted twice.  The red curve is the
best-fit model for the radial velocity variation of the star after the
subtraction of the effect of \planetb, \planetd, \koie, and \koif. {\it Lower Panel:} Same as above, but showing the radial
velocities (in gray, with binned points in blue) and model (in red) of \starname\ after correcting for effect of \planetc, \planetd, \koie, and \koif, leaving the effect of only \planetb\ and plotted as a function of its orbital phase.}
\label{fig:rv_phased}
\end{center}
\end{figure}

We adopted the median of the posterior distribution of each parameter as our estimate of its value, and we assigned the uncertainties by 
identifying the adjacent ranges of each parameter that encompassed 34\% of the values above and below the median.  We estimated the parameter 
distributions for the planetary masses and radii by combining the stellar mass and radius distributions from the evolution track fits 
with the model parameter distributions.  The parameter estimates for the star are listed in Table~\ref{tab:s-params}, and the parameter 
estimates for the confirmed planets \planetb, \planetc, and \planetd\ are listed in Table~\ref{tab:p-params}.  The light curves phased to 
the times of transit of each planet or candidate are shown in Figure~\ref{fig:lc_phased}. In Figure~\ref{fig:rv_phased} we show 
the radial velocities phased to the orbital periods of \planetb\ and \planetc.

We note that the values of $a/R_{\star}$ for K00070.04 and K00070.05 were misstated in Table~1 of \citet{fressin12}. The correct values 
are $14.4^{+1.4}_{-1.2}$ and $31.3^{+3.0}_{-2.5}$, respectively. The values of the stellar parameters were also stated incorrectly in the 
same table, and should read $T_{\rm eff} = 5455 \pm 100$\,K, $log g = 4.4 \pm 0.1$, ${\rm [Fe/H]}= +0.01 \pm 0.04$, 
$v \sin i < 2$~km~s$^{-1}$, and $L/L_{\sun} =0.71_{-0.29}^{+0.14}$. These changes have no effect on the conclusions of \citet{fressin12}.

\begin{deluxetable}{lcc}
\tabletypesize{\scriptsize}
\tablewidth{0pc}
\tablecaption{Parameters for the Star \starname. \label{tab:s-params}}
\tablehead{\colhead{Parameter}	& \colhead{Value} 	& \colhead{Notes}}
\startdata
Right Ascension (J2000)			& \kicra 			& \\
Declination (J2000)				& \kicdec 			& \\
\kepler\ Magnitude				& \kepmag 		& \\
r Magnitude					& \rmag			& \\
\sidehead{\em Spectroscopically Determined Parameters}
Effective temperature \teff~(K)		& $5455 \pm 100$	& A \\
Spectroscopic gravity \logg~(cgs)    & $4.4 \pm 0.1$	        & A \\
Metallicity \feh				        & $0.01 \pm 0.04$	& A \\
Mt.~Wilson S-value				& $0.183 \pm 0.005$& A \\
\logrhk						& $-4.93 \pm 0.05$	& A \\
Projected rotation \vsini~(\kms)		& $< 2$	                & A \\
Mean radial velocity~(\kms)		& $-21.87 \pm 0.96$	& A \\
Radial Velocity Instrumental Zero Point $\gamma$~(\ms)	& $-3.54^{+0.68}_{-1.02}$ & B \\

\sidehead{\em Derived stellar properties}
Mass \mstar~(\msun)			& $0.912 \pm 0.034$		& C \\
Radius \rstar~(\rsun)			& $0.944^{+0.060}_{-0.095}$	& C \\
Density \rhostar~(cgs)                      & $1.51 \pm 0.38$                    & C \\  
Luminosity \lstar~(\lsun)			& $0.71^{+0.14}_{-0.29}$	& C \\
Age~(Gyr)                                        & $8.8^{+4.7}_{-2.7}$               & C \\
Distance~(pc)					& $290 \pm 30$		        & C \\ 
\enddata
\tablecomments{
A: Based on the spectroscopic analysis (\S\ref{sec:spectroscopy}).\\
B: Not a physical parameter, reported here for completeness.\\
C: Based on a comparison of stellar evolutionary tracks to constraints from the spectroscopically-determined parameters (\S\ref{sec:spectroscopy}) and the transit durations (\S\ref{sec:model}).\\
}
\end{deluxetable}

\begin{deluxetable}{lcccc}
\tabletypesize{\scriptsize}
\tablewidth{0pc}
\tablecaption{Physical and orbital parameters for \planetb , \planetc , and \planetd\ \label{tab:p-params}}
\tablehead{\colhead{Parameter}			&  \colhead{\planetb}                                       &  \colhead{\planetc}                     &  \colhead{\planetd}                               & \colhead{Notes}}
\startdata
Orbital period $P$ (days)					& $3.6961219^{+0.0000043}_{-0.0000064}$ & $10.854092 \pm 0.000013$       & $77.61184^{+0.00015}_{-0.00037}$       & A \\
Midtransit time $T_0$ (BJD)				& $2454967.50027^{+0.00058}_{-0.00068}$ & $2454971.60758 \pm 0.00046$ & $2454997.7271^{+0.0016}_{-0.0019}$   & A \\
Scaled semi-major axis $a/\rstar$		        & $10.29^{+0.97}_{-0.83}$ 			      & $21.1^{+2.0}_{-1.7}$                  & $78.3^{+7.4}_{-6.3}$ 			             & A \\
Scaled planet radius $R_p/\rstar$		        & $0.01855^{+0.00026}_{-0.00031}$ 	      & $0.02975 \pm 0.00032$             & $0.02670^{+0.00046}_{-0.00069}$         & A \\
Impact parameter $b$                                      & $0.633^{+0.025}_{-0.021}$ 			      & $0.594^{+0.018}_{-0.021}$        & $0.588^{+0.041}_{-0.032}$ 		     & A \\
$e \, cos ({\omega})$				        & $-0.004^{+0.033}_{-0.055}$ 	                      & $-0.097^{+0.054}_{-0.072}$       & $-0.002^{+0.025}_{-0.045}$ 	             & A \\      
$e \, sin ({\omega})$	                                         & $-0.021^{+0.021}_{-0.030}$                        & $-0.011^{+0.031}_{-0.022}$      & $-0.007^{+0.025}_{-0.022}      $ 		     & A \\     
Orbital inclination $i$ (deg)				& $86.50^{+0.36}_{-0.31}$			      & $88.39^{+0.16}_{-0.14}$            & $89.570^{+0.043}_{-0.048}     $		     & A \\
Orbital eccentricity $e$					& $< 0.32$ 						      & $< 0.40$                                     & $< 0.60$ 						     & A \\
Orbital semi-amplitude $K$ (\ms)                     & $3.72^{+0.76}_{-1.09}$ 			      & $4.83^{+1.03}_{-0.98}$              & $1.2^{+1.0}_{-1.3}$ 				     & A \\ 
Mass $M_p$ $(M_{\Earth}$)				& $8.7^{+2.1}_{-2.2}$ 				      & $16.1^{+3.3}_{-3.7}$                  & $< 20.1~(2~\sigma)$				     & B \\
Radius $R_p$ $(R_{\Earth}$)				& $1.91^{+0.12}_{-0.21}$ 		  	      & $3.07^{+0.20}_{-0.31}$              & $2.75^{+0.17}_{-0.30}$  			     & B \\
Density ${\rho}_p$ (\gcmc)				& $6.5^{+2.0}_{-2.7}        $			      & $2.91^{+0.85}_{-1.08}$              & $< 4.07$						     & B \\
Orbital semi-major axis $a$ (AU)			& $0.04537^{+0.00054}_{-0.00060}$ 	      & $0.0930 \pm 0.0011$                 & $0.3453^{+0.0041}_{-0.0046}$ 		     & B\\
Equilibrium temperature \teq\ (K)	                & $1014$ 							      & $713$                                         & $369$ 						     & C \\
\enddata
\tablecomments{
A: Based on the joint modeling (\S\ref{sec:model}) of the light curve and radial velocities, with eccentricities constrained to avoid orbit crossing.\\
B: Based on the results from A and estimates of \mstar\ and/or \rstar\ from Table~\ref{tab:s-params}.\\
C: Calculated assuming a Bond albedo of 0.5 and isotropic re-radiation of absorbed flux from the entire planetary surface.\\
}
\end{deluxetable}

\clearpage

\subsection{Limits on Motion of Photocentroid}
\label{sec:centroid}

While the analysis above provides the parameter estimates of the five planet candidates under the assumption
that each are planets orbiting \starname, it does not address the concern that some or all of these five candidates
result instead from an astrophysical false positive (i.e. a blend of several stars within the \kepler\ photometric
aperture, containing an eclipsing component).  In \S\ref{sec:blender} we use the {\tt BLENDER} method to demonstrate 
that this possibility is extremely unlikely for \planetb, \planetc, and \planetd, and it is this {\tt BLENDER} work
that is the basis for our claim that each of these three objects are planets.  Another means
to identify astrophysical false positives is to examine the \kepler\ pixel data to detect the shift in the photocentroid of the image 
\citep[e.g.][]{batalha10, torres11, ballard11} of \starname\ during times of transit, which we discuss below. 
Although we do not use the results presented below as part of the {\tt BLENDER} work, we
include a description of it here as it provides an independent argument against the hypothesis that the photometric signals
result from an astrophysical false positive and not from planetary companions to \starname.

We use two methods to examine the \kepler\ pixel data to evaluate the location of the photocenter and thus to search for 
astrophysical false positives: (1) the direct measurement of the source location via difference images, the PRF centroid method,
and (2) the inference of the source location from photocenter motion associated with the transits, the flux weighted centroid method.
In principle both techniques are similarly accurate, but in practice the flux weighted centroid technique is more sensitive to noise for 
low signal-to-noise ratio (SNR) transits.  We use both techniques because they are
both subject to biases due to various systematics, but respond to those systematics in different ways.

In our difference image analysis \citep{torres11}, we evaluate the difference between average in-transit pixel images and 
average out-of-transit images.  In the absence of pixel-level systematics, the pixels with the highest flux in the 
difference image will form a star image at the location of the transiting object, with amplitude equal to the depth of the transit.
A fit of the \kepler\ pixel response function \citep[PRF,][]{bryson10} to both the difference and out-of-transit
images provides the offset of the transit source from \starname. We measure difference images separately in each
quarter, and estimate the transit source location as the robust uncertainty-weighted average of the quarterly results.

We measure photocenter motion by computing the flux-weighted centroid of all pixels downlinked for \starname,
generating a centroid time series for row and column.  We fit the modeled transit to the whitened centroid time series
transformed into sky coordinates.  We perform a fit for each quarter, and infer the source location 
by scaling the difference of these two centroids by the inverse of the flux as described in \cite{jenkins10c}.  

Both the difference image and photocenter motion methods are vulnerable to various systematics, which
may bias the result.  The PRF fit to the difference and out-of-transit pixel images is biased by PRF errors
described in \cite{bryson10}.  The photocenter technique is biased by stars not being completely captured by 
the available pixels.  These types of biases will vary from quarter to quarter.  Both methods are vulnerable to crowding,
depending on which pixels are downlinked, which varies from quarter to quarter.  We ameliorate these biases by taking the 
uncertainty-weighted robust average of the source locations over available quarters.  Because the biases of 
these difference image and photocenter motion techniques differ, we take agreement of the multi-quarter averages 
as evidence of that we have faithfully measured the source location of the transit signal.

\begin{figure}
\begin{center}
\includegraphics[scale = 0.35]{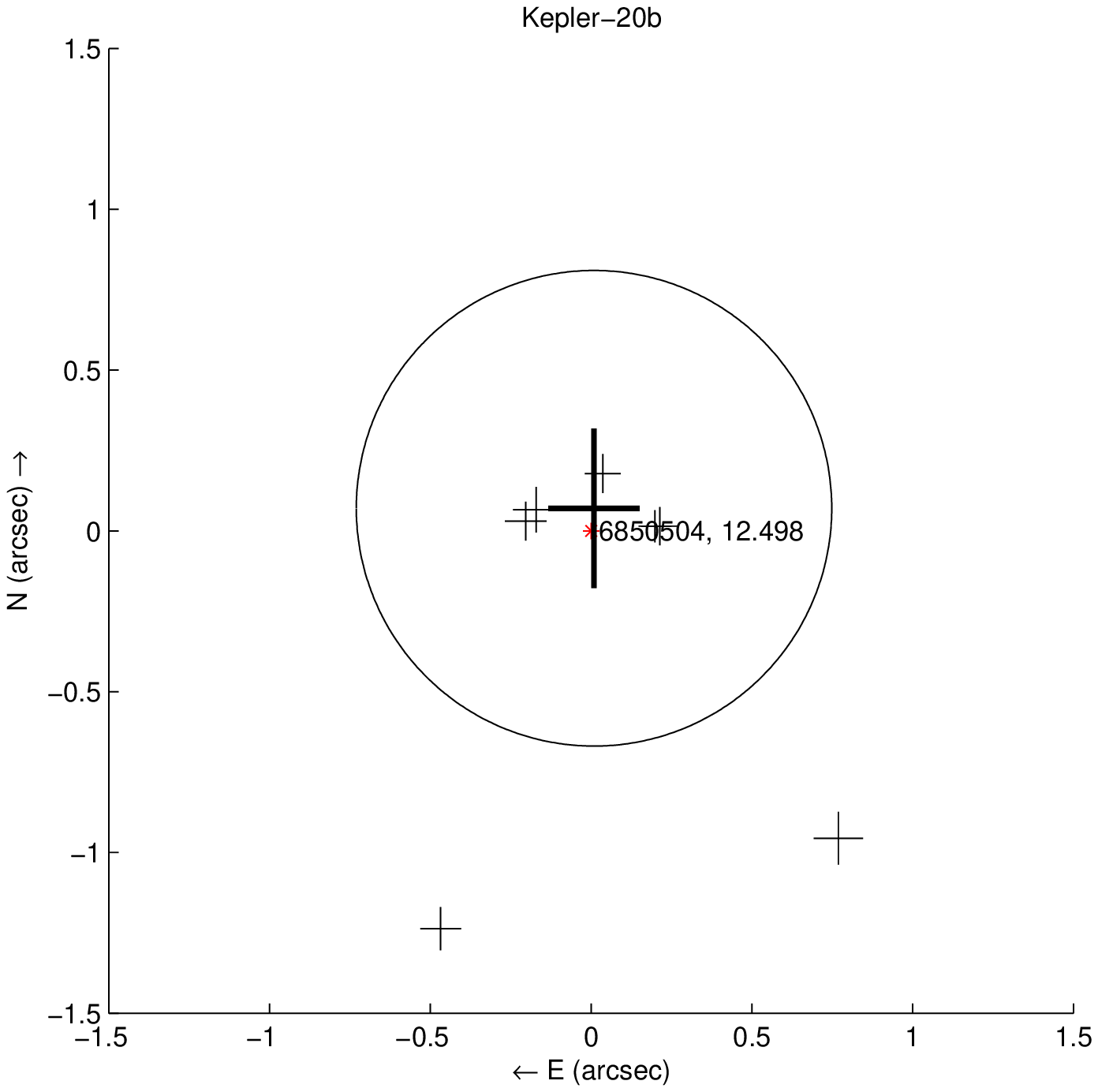}
\includegraphics[scale = 0.35]{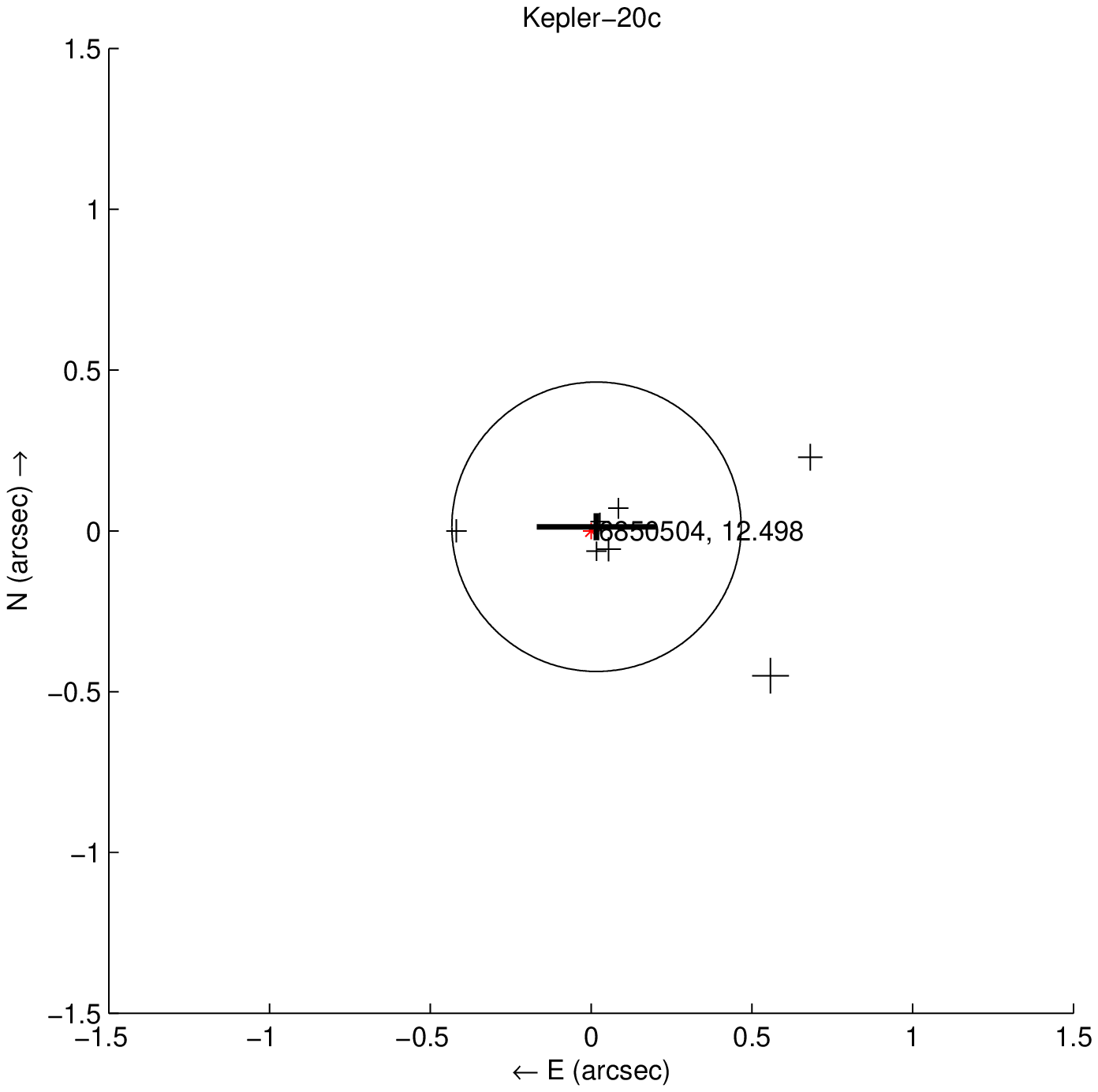}
\includegraphics[scale = 0.35]{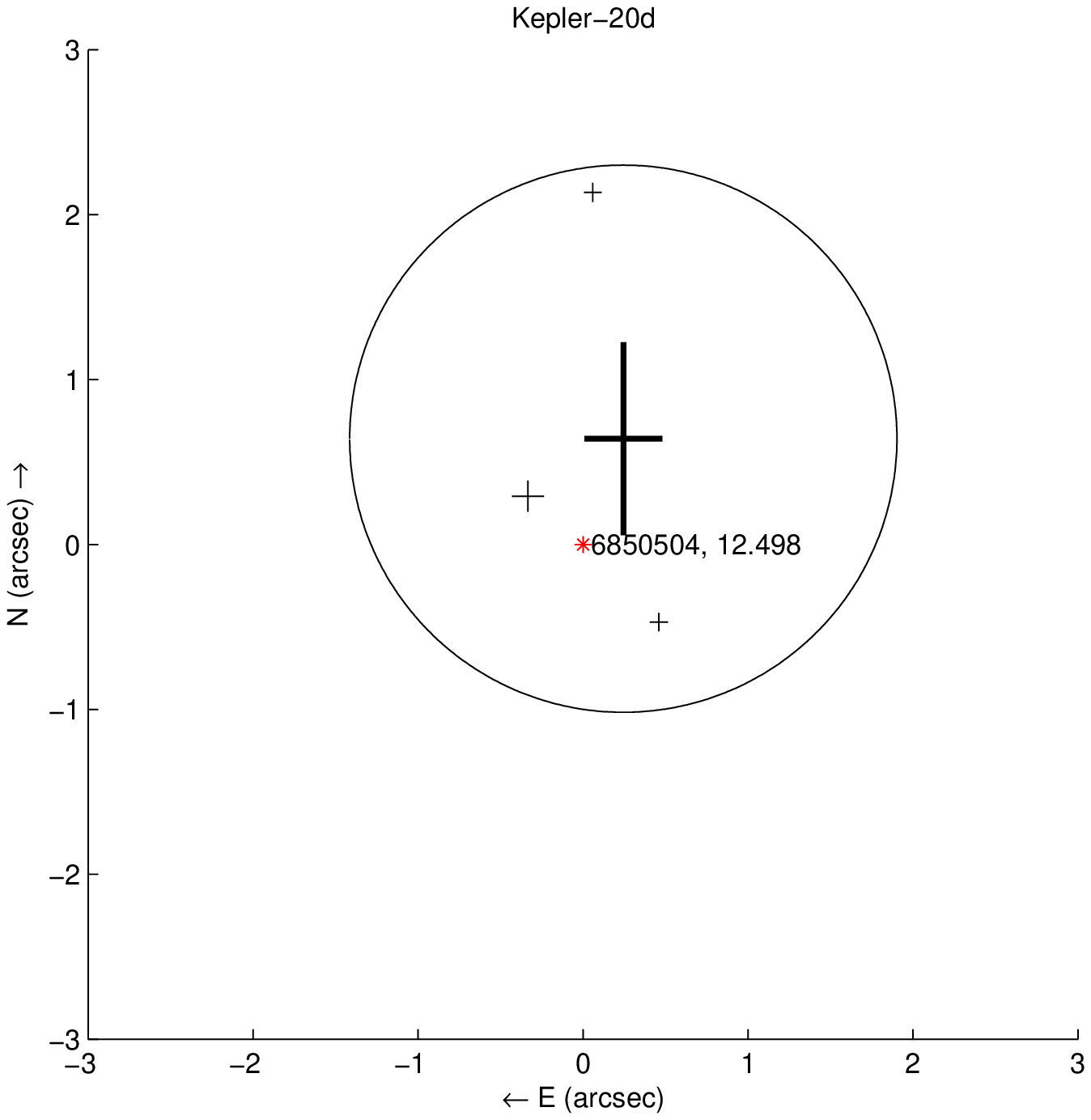}
\includegraphics[scale = 0.35]{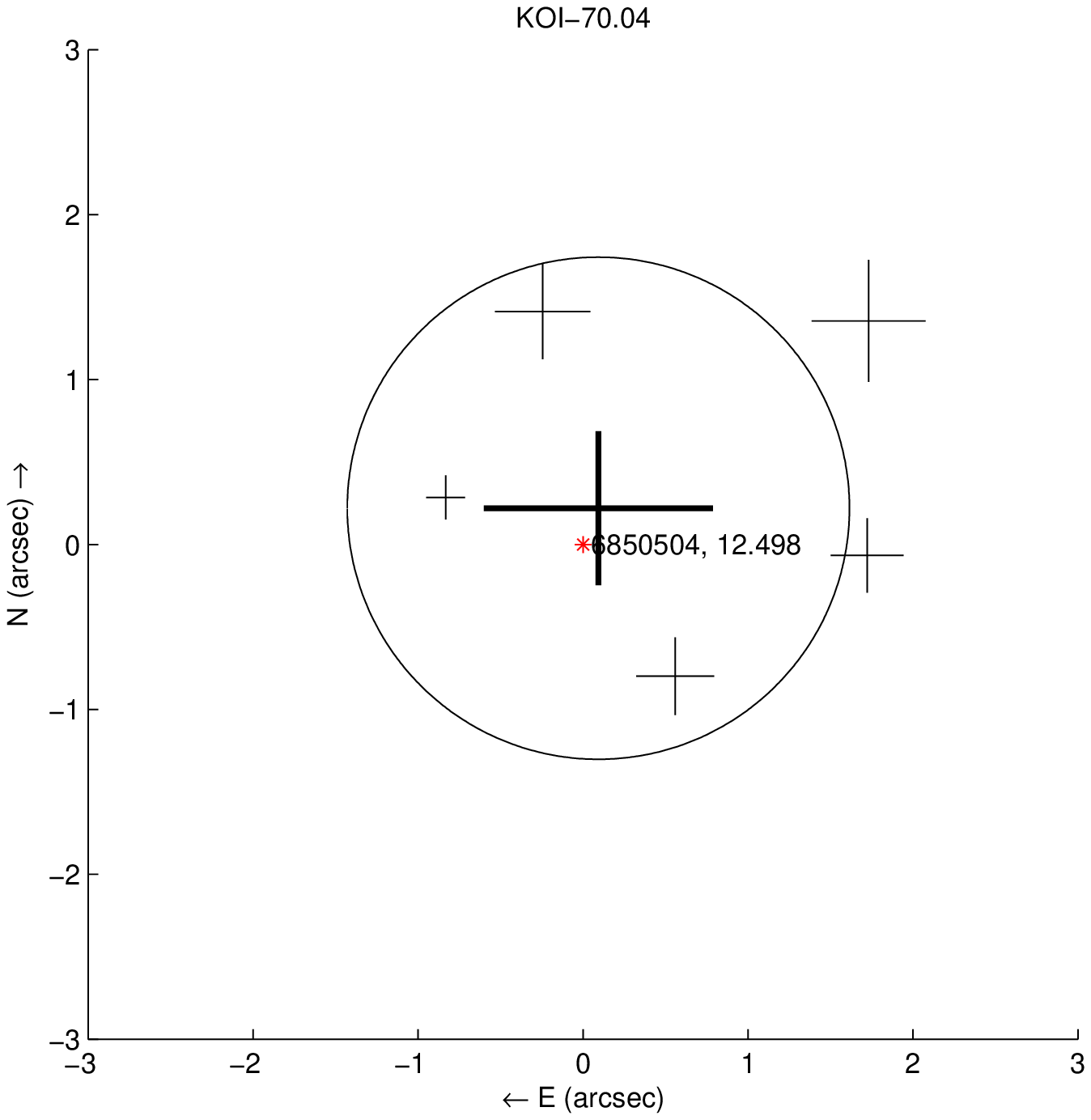}
\includegraphics[scale = 0.35]{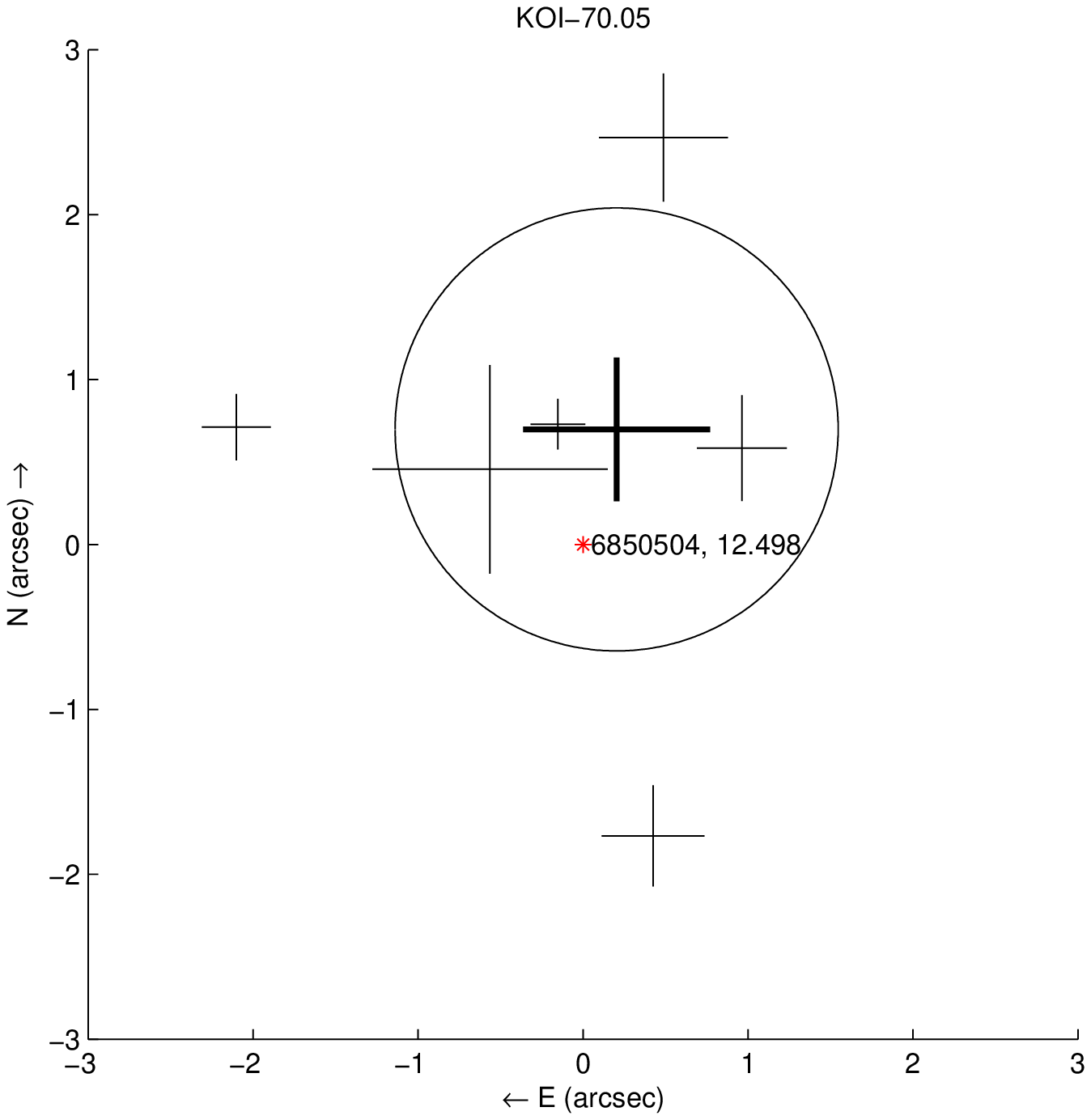}
\caption{Quarterly and average source locations using the difference image (PRF centroid) method for \planetb\ (top row left), 
\planetc\ (top row center), \planetd\ (top row right), \koie\ (bottom row left), and \koif\ (bottom row right).
The light crosses show the individual quarter measurements and the heavy crosses show the 
uncertainty-weighted robust average.  The length of the crosses show the $1~\sigma$ uncertainty of each 
measurement in RA and Dec.  The circles have radius $3~\sigma$ and are centered at the average measured source 
location.  The location of \starname\ is shown by the red asterisk and labeled with its KIC number.  
In all the panels, the offset between the measured source location and the target is less than 2~$\sigma$.}
\label{fig:planet_offsets}
\end{center}
\end{figure}

Table \ref{tab:planet_offsets} provides the offsets of the transit signal source from \starname\ averaged over
Q1$-$Q7 for all five planet candidates.  The quarterly measurements and averages for the PRF centroid method are shown in Figure~\ref{fig:planet_offsets}.  
All the average offsets are within 2 sigma of \starname.

\begin{deluxetable}{lcccc}
\tabletypesize{\scriptsize}
\tablewidth{0pc}
\tablecaption{Offsets between Photocenter of Transit Signal and \starname}
\tablehead{\colhead{Candidate}	&  \colhead{PRF Centroid [{\it arcsec}]} &  \colhead{Significance\tablenotemark{a}} & \colhead{Flux-weighted Centroid [{\it arcsec}]} & \colhead{Significance\tablenotemark{a}}}
\startdata
\planetb &   $0.071\pm0.25$  & 0.29 &  $0.41\pm0.24$  &  1.72   \\
\planetc &   $0.021\pm0.17$  & 0.12 &  $0.072\pm0.22$ & 0.32    \\
\planetd &   $0.69\pm0.74$    & 0.92 &  $3.07\pm1.96$  &  1.59   \\
\koie      &   $0.24\pm0.51$    & 0.47 &  $2.14\pm1.79$  &  1.20   \\
\koif       &   $0.73\pm0.45$    & 1.62 &  $1.76\pm1.91$  &  0.93   \\
\enddata
\tablenotetext{a}{offset/uncertainty}
\label{tab:planet_offsets}
\end{deluxetable}

\clearpage

\subsection{Stellar Activity and Rotation from the \kepler\ Light Curve}
\begin{figure}
\begin{center}
\includegraphics[scale=0.75]{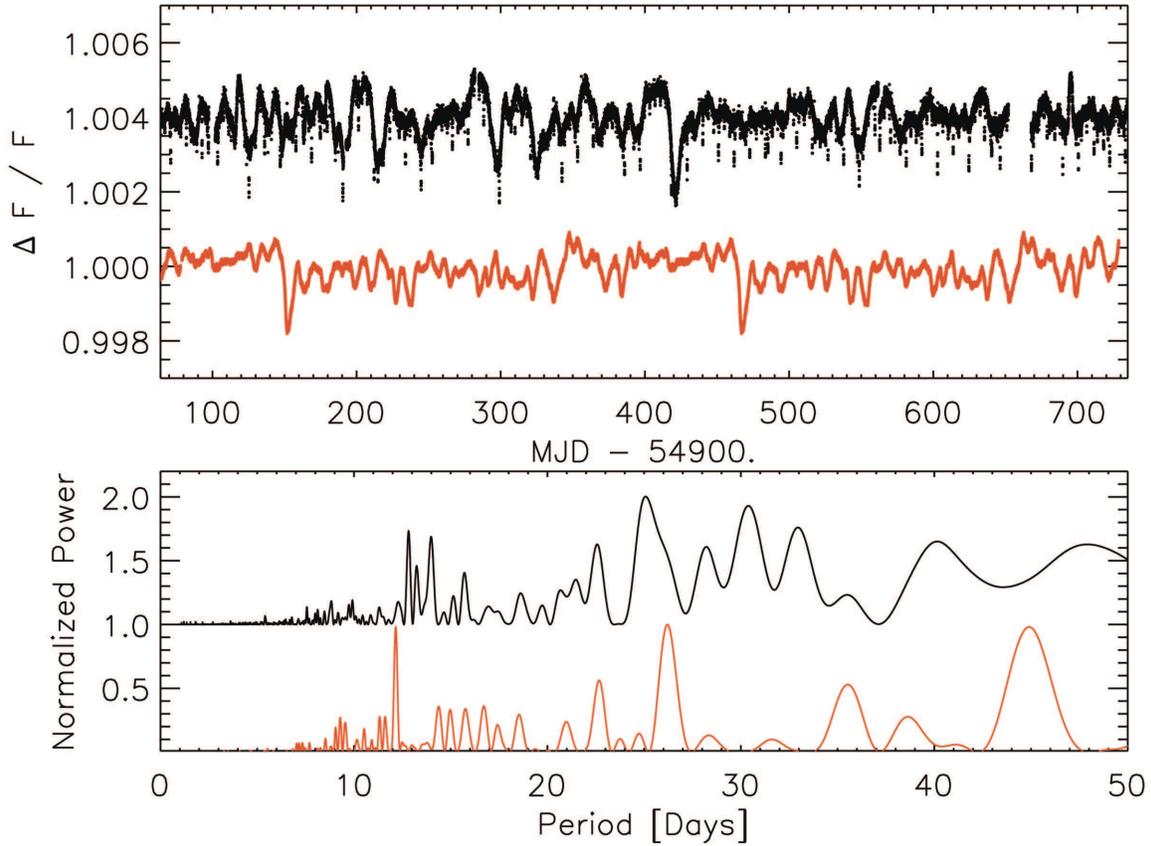}\\
\caption{{\it Upper panel:} PDC-MAP corrected lightcurve for \starname\ from Q1 to Q8 (black points), showing \starname's intrinsic
stellar variability after removal of instrumental effects. Orange points show the lightcurve for
the 2001 active sun from the SOHO Virgo instrument $g+r$ observations for comparison \citep[lightcurve prepared
as described in ][]{basri10}.  {\it Lower panel:} Lomb-Scargle periodogram for each of the two curves appearing in the upper panel.}
\label{fig:lc_variability}
\end{center}
\end{figure}

While the polynomial-fitting
approach in \S\ref{sec:model} is well-suited to cleaning the time series for the transit analysis, 
it annihilates any long-term variability, such as that due to rotationally modulated star spots.
Subsequent to the analysis in \S\ref{sec:model}, we obtained a version of the \kepler\ photometry
from Q1$-$Q8 using the new pipeline {\tt PDC-MAP} \citep{jenkins11}, which more effectively removes non-astrophysical systematics in
the photometry while leaving the stellar variability intact. We used this {\tt PDC-MAP} corrected 
light curve to evaluate the rotational period and stellar activity of the star. We computed a Lomb-Scargle periodogram, 
and found the highest power peak at a period of 25 days, with a lobe on that peak at around 26 days. This 
peak is also accompanied by significant power at periods between 24 and 32 days. The distribution of periodicities and in
particular the lobed, broad appearance of the peak with the highest power is strongly reminiscent
of the activity behavior of the Sun, where differential rotation is responsible for a range of
periods from approximately 25 days at the equator to 34 days at the poles. Indeed, the amplitude of
spot-related variability on \starname\ is very similar to that of the active Sun, as measured during
the 2001 season by the SOHO Virgo instrument. Using the SOHO light curves treated to resemble
Kepler photometry \cite[as described in][]{basri11}, we compared the amplitude of variability
of \starname\ and the active Sun; the two light curves (and the
Lomb-Scargle periodogram of each) are shown in Figure~\ref{fig:lc_variability}.
We found that \starname\ has spot modulation roughly 30\% higher in
amplitude than that of our Sun. Our estimate of the rotation period (above) for \starname\ is consistent with both 
its spectroscopically estimated \vsini~$<2$~\kms\ and an estimate of the rotation period, 31~d, based on its Ca~H and K 
emission \logrhk\ (see \S\ref{sec:spectroscopy}). Together, the period and variability indicate that \starname\ 
is very similar to (but perhaps somewhat more active than) our own star. 

\clearpage

\section{Follow-Up Observations}

\subsection{High-Resolution Imaging}

In order to place limits upon the presence of stars near the target that could be the source
of one or more of the transit signals, we gathered high-resolution imaging of \starname\ with three separate facilities:
We obtained near-infrared adaptive optics images with both the Palomar
Hale 5m telescope and the Lick Shane 3m telescope,
and we gathered optical speckle observations with the Wisconsin Indiana Yale NOAO (WIYN) 3.5m telescope.  Ultimately we used only the
Palomar observations in our {\tt BLENDER} analysis (\S4) as these were the most constraining, but we describe all three sets of observations here for completeness.

\subsubsection{Adaptive Optics Imaging}\label{sec:ao}
We obtained near-infrared adaptive optics imaging of \starname\ on the night
of UT 2009 September 09 with the Palomar Hale 5m telescope and the PHARO
near-infrared camera \citep{hayward2001} behind the Palomar adaptive optics
system \citep{troy00}.  We used PHARO, a $1024\times1024$ HgCdTe infrared array,
in 25.1 mas/pixel mode yielding a field of view of
$25^{\prime\prime}$.  We gathered our observations in the $J$ ($\lambda_0 =
1.25\mu$m) filter.  We collected the data in a standard 5-point quincunx
dither pattern of $5^{\prime\prime}$ steps interlaced with an off-source
($60^{\prime\prime}$ East and West) sky dither pattern. The integration
time per source was 4.2 seconds at $J$.  We acquired a total of 15 frames 
for a total on-source integration time of 63 seconds.  The adaptive
optics system guided on the primary target itself; the full width at half maximum (FWHM)
of the central core of the resulting point spread function was $0\farcs07$.  

\begin{figure}
\includegraphics[angle=90,scale=0.7,keepaspectratio=true]{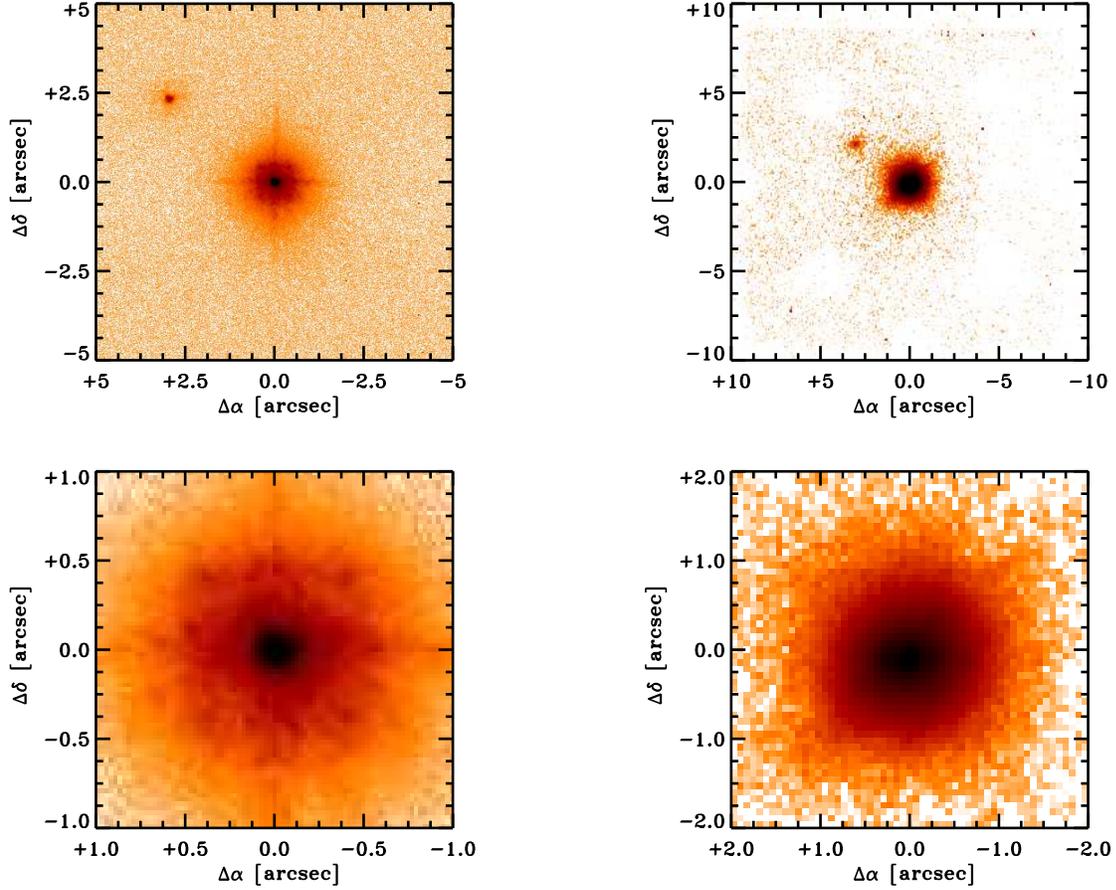}
\figcaption{Palomar $J$ (left) and Lick $K_{s}$ (right) adaptive
optics images of \starname.  The top row displays a $10\arcsec \times
10\arcsec$ field of view for the Palomar $J$ image and a $20\arcsec \times
20\arcsec$ for the Lick $K_s$ image.  The bottom row displays zoomed images
highlighting the area immediately around the target . \label{fig:paloAO}}
\end{figure}

We further obtained near-infrared adaptive optics imaging on the night of UT 2011
June 17 with the Lick Shane 3m telescope and the IRCAL 
near-infrared camera \citep{lloyd00} behind the natural guide star adaptive
optics system.  IRCAL is $256\times256$ pixel PICNIC array with a plate
scale of 75.6 mas/pixel, yielding a total field of view of
$19\farcs6$.  We gathered our observations using the $K_s$
($\lambda_0 = 2.145\mu$m) filter, and, as with the Palomar observations,
we used a standard 5-point dither pattern. The integration time per
frame was 120 seconds; we acquired 10 frames for a total on-source
integration time of 1200 seconds.  The adaptive optics system guided on the
primary target itself; the FWHM of the central core of the resulting point
spread function were $0.79^{\prime\prime}$.   The final coadded
images at $J$ and $K_s$ are shown in Figure~\ref{fig:paloAO}.

In addition to \starname, we detected two additional sources.  The
first source is $3\farcs8$ to the northeast of the target and is
fainter by $\Delta J \approx 4.5$ magnitudes and  $\Delta Ks \approx 4.2$
magnitudes. The star has an infrared color of $J-K_{s} = 0.19\pm0.02$ mag which
yields an expected \kepler\ magnitude of $Kp = 16.1 \pm 0.2$ mag
\citep{howell11a}. A much fainter source, at $11^{\prime\prime}$ to the
southeast, was detected only in the Palomar $J$ data and is  $\Delta J =
8.5$ magnitudes fainter than the primary target.  The fainter source, based
upon expected stellar $Kp-J$ colors \citep{howell11a}, has an expected
\kepler\ magnitude of $Kp=21.0 \pm 0.7$ mag. 

\begin{figure}
\includegraphics[angle=0,scale=0.47,keepaspectratio=true]{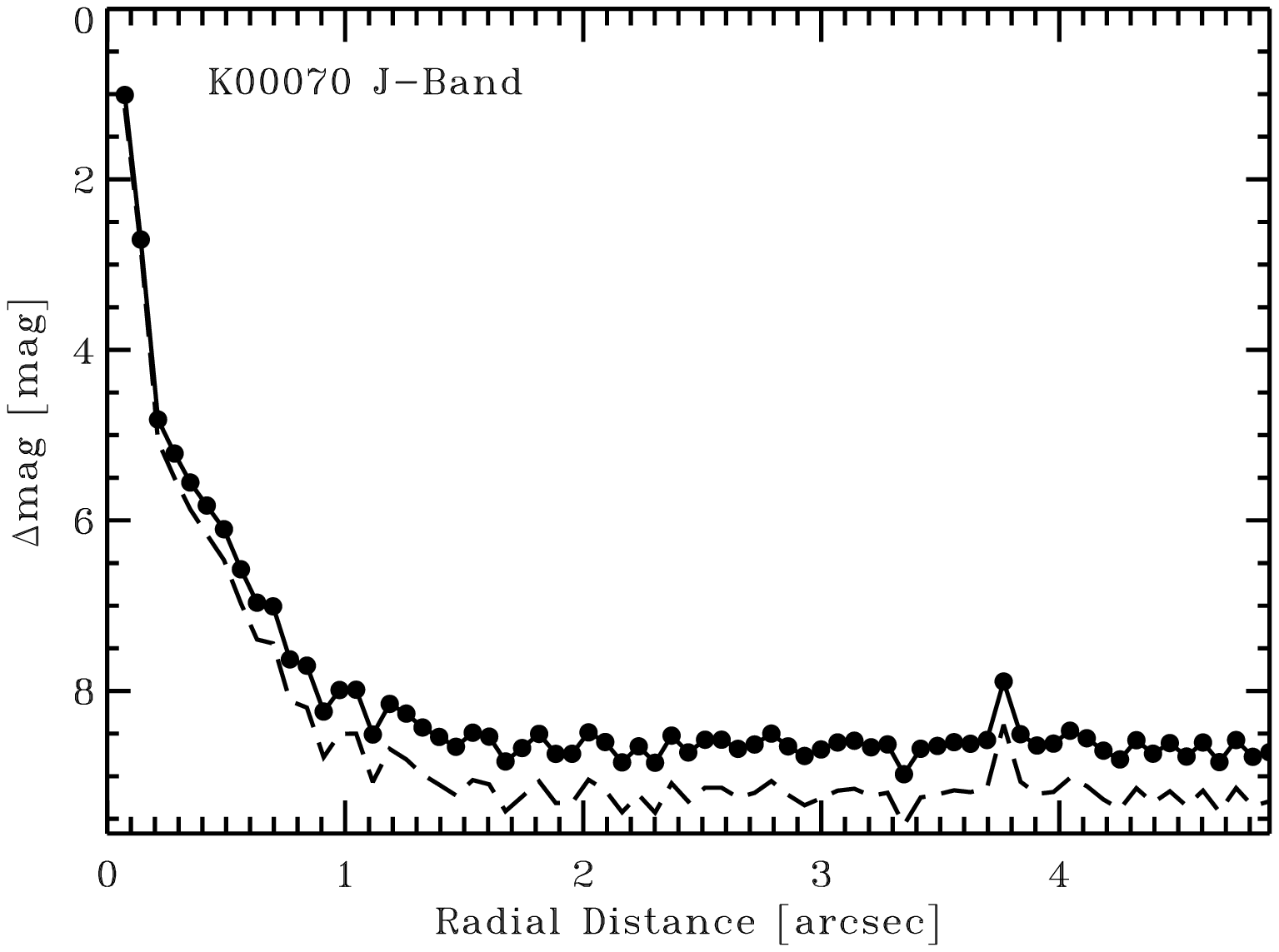}
\includegraphics[angle=0,scale=0.47,keepaspectratio=true]{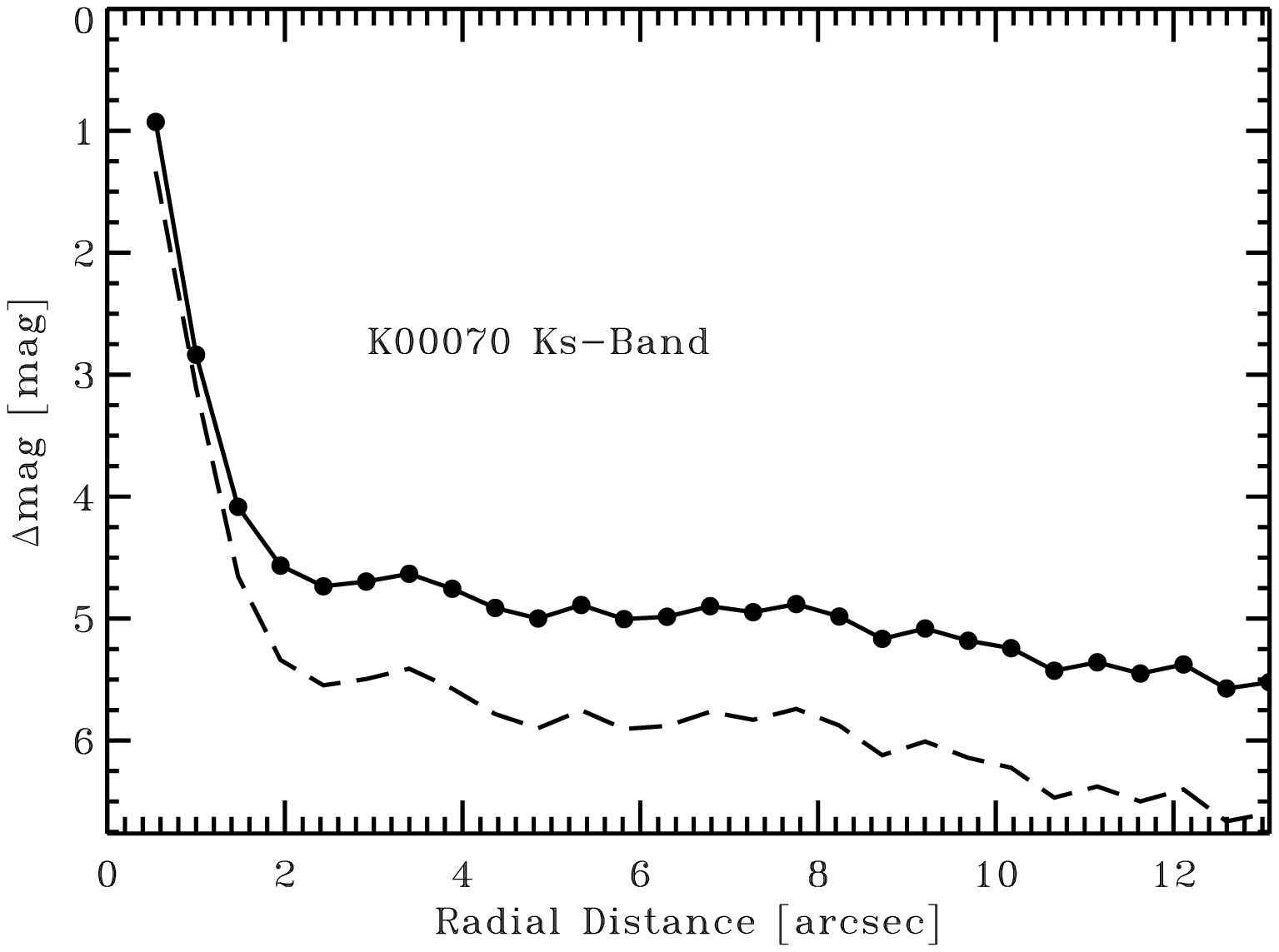}
\figcaption{{\it Left panel:} The sensitivity limits of the Palomar $J$-band adaptive optics imaging are
plotted as a function of radial distance from the primary target.  The
filled circles and solid line represent the measured $J$-band limits; each
filled circle represents one step in FWHM.  The dashed line represents 
the derived corresponding limits in the \kepler\ bandpass based upon the
expected $Kp - J$ colors \citep{howell11a}.  {\it Right panel:} Same as above,
but showing the sensitivity limits of the Lick 
$K_{s}$-band adaptive optics imaging.  The dashed line is based upon the
expected $Kp - K_{s}$ colors.\label{fig:paloAO_Jlim}} 
\end{figure}

No additional sources were detected in the imaging.  We determined the sensitivity 
limits of the imaging by calculating the noise in concentric rings
radiating out from the centroid position of the primary target, starting at
one FWHM from the target with each ring stepped one FWHM from the previous
ring.  The $3~\sigma$ limits of the $J$-band and $K$-band imaging were approximately
20 mag and 16 mag, respectively (see Figure~\ref{fig:paloAO_Jlim}).  The respective $J$-band and $K$-band imaging limits
are approximately $8.5$ and $5.5$ mag fainter than the target, corresponding to
contrasts in the \kepler\ bandpass of approximately $9$ mag and $6.5$ mag.

\subsubsection{Speckle Imaging}
\label{sec:speckle}
We obtained speckle observations of \starname\ at the WIYN 3.5m telescope on two different nights, UT 2010 June 18 and
UT 2010 September 17. We gathered both sets of observations with the new dual channel
speckle camera described in \citet{horch11}. On both nights the data consisted of 3 sets of 1000 exposures
each with an individual exposure time of 40ms, with images gathered simultaneously in two filters. The data collection, 
reduction, and image reconstruction process are described in the aforementioned
paper as well as in \citet{howell11b}, and the latter presents details of the
2010 observing season of observations
with the dual channel speckle camera for the \kepler\ follow up program.
 
On both occasions our speckle imaging did not detect a companion to \starname\ within an annulus
of $0.05-1.8$ arcsec from the target. The September observation yielded
detection limits of 3.82 (in $V$) and 3.54 (in $R$)
magnitudes fainter than \starname. The June observation
yielded limits of 3.14 and 4.92 fainter in $V$ and $R$ respectively. Therefore we rule out the
presence of a second star down to 3.82~magnitudes fainter in $V$ and 4.92~magnitudes fainter in $R$ 
over an angular distance of 0.05$-$1.8 arcseconds from \starname. 

\clearpage

\subsection{Photometry with the {\it Spitzer}\ Space Telescope}
\label{sec:spitzer}

An essential difference between true planetary transits and astrophysical false positives resulting from
blends of stars is that the depth of a planetary transit is achromatic
(neglecting the small effect of stellar limb-darkening), whereas false positives are not (except in the unlikely case that
the effective temperatures of the contributing stars are extremely similar).  By providing infrared
time series spanning times of transit, the \wspitzer\ Mission has assisted in the validation of many transiting planet
systems, including Kepler-10 \citep{fressin11a}, Kepler-14 \citep{buchhave11}, Kepler-18 \citep{cochran11}, 
Kepler-19 \citep{ballard11}, and CoRoT-7 \citep{fressin11b}.  We describe below our observations and analysis
of \wspitzer\ data spanning transits of \planetc\ and \planetd, which provide independent support of their planetary nature.

\subsubsection{Observations and Extraction of the \wspitzer\ Time Series}\label{spitzer_obs}

We used the IRAC camera \citep{fazio04} on board the \spitzer\ Space Telescope \citep{werner04} to observe \starname\ 
spanning one transit of \planetc\ and two transits of \planetd.  We gathered our observations at 4.5~\micron\ as
part of program ID 60028. The visits lasted 8.5~hours for \planetc\ and 16.5~hours for both visits of \planetd. 
We used the full-frame mode ($256\times256$ pixels) with an exposure time of 12~s per image, which yielded 2451 and 4643 images 
per visit for \planetc\ and \planetd, respectively. 

The method we used to produce photometric time series from the images is described in \cite{desert09}.
It consists of finding the centroid position of the stellar point spread function (PSF) and performing aperture photometry using a circular aperture 
on individual Basic Calibrated Data (BCD) images delivered by the \emph{Spitzer} archive.
These files are corrected for dark current, flat-fielding, detector non-linearity and converted into flux units.  
We converted the pixel intensities to electrons using the information given in the detector gain and exposure time provided in the image headers; 
this facilitates the evaluation of the photometric errors. We adjusted the size of the photometric aperture to yield the smallest errors; for these data
the optimal aperture was found to have a radius of $3.0$~pixels. We found that the transit depths and errors varied only weakly with the 
aperture radius for each of the light curves. We used a sliding median filter to identify and trim outliers that differed in flux or positions by greater than $5~\sigma$. 
We also discarded the first half-hour of observations, which are affected by a significant jitter before the telescope stabilizes.
We estimated the background by fitting a Gaussian to the central region of the histogram of counts from the full array. 
Telescope pointing drift resulted in fluctuations of the stellar centroid position, which, in
combination with intra-pixel sensitivity variations, produces systematic noise in the raw light curves.  
A description of this effect, known as the pixel-phase effect, is given in the \spitzer/IRAC data handbook \citep{reach06} 
and is well known in exoplanetary studies (e.g. \citealt{charbonneau05,knutson08}). After correction for this effect (see below) we found 
that the point-to-point scatter in the light curve indicated an achieved SNR of $220$ per image, corresponding to 85\% of the theoretical limit.

\subsubsection{Analysis of the \wspitzer\ Light Curves}\label{spitzer_model}

We modeled the time series using a model that was a product of two functions, one describing the transit shape and the other describing 
the variation of the detector sensitivity with time and sub-pixel position, as described in \cite{desert11a}. 
For the transit light curve model, we used the transit routine \texttt{OCCULTNL} from \cite{mandel02}.
This function depends on the parameters $(R_p / R_{\star})_i$, $(a / R_{\star})_i$, $b_i$, and $T_{0,i}$, where $i=\{$\planetc, \planetd$\}$, the two candidate planets
for which we gathered observations.  The contribution of stellar limb-darkening at 4.5~\micron\ is negligible given the low precision of our \wspitzer\ data 
and so we neglect this effect. We allow only $(R_p / R_\star)_i$ to vary in our analysis; the other parameters are set to the values derived from 
the analysis of the \kepler\ light curve (see Table~\ref{tab:p-params}). Because of the possibility of transit-timing variations (see \S5), we set the values of
$T_{0,i}$ to the values measured from \kepler\ for the particular event.
Our model for the variation of the instrument response consists of a sum of a linear function of time and a quadratic function (with four parameters) of the 
$x$ and $y$ sub-pixel image position. We simultaneously fit the instrumental function and the transit shape for each individual visit.
The errors on each photometric point were assumed to be identical, and were set to the root-mean-squared residuals to the initial best fit obtained.

\begin{figure}
\begin{center}
 \includegraphics[width=4in]{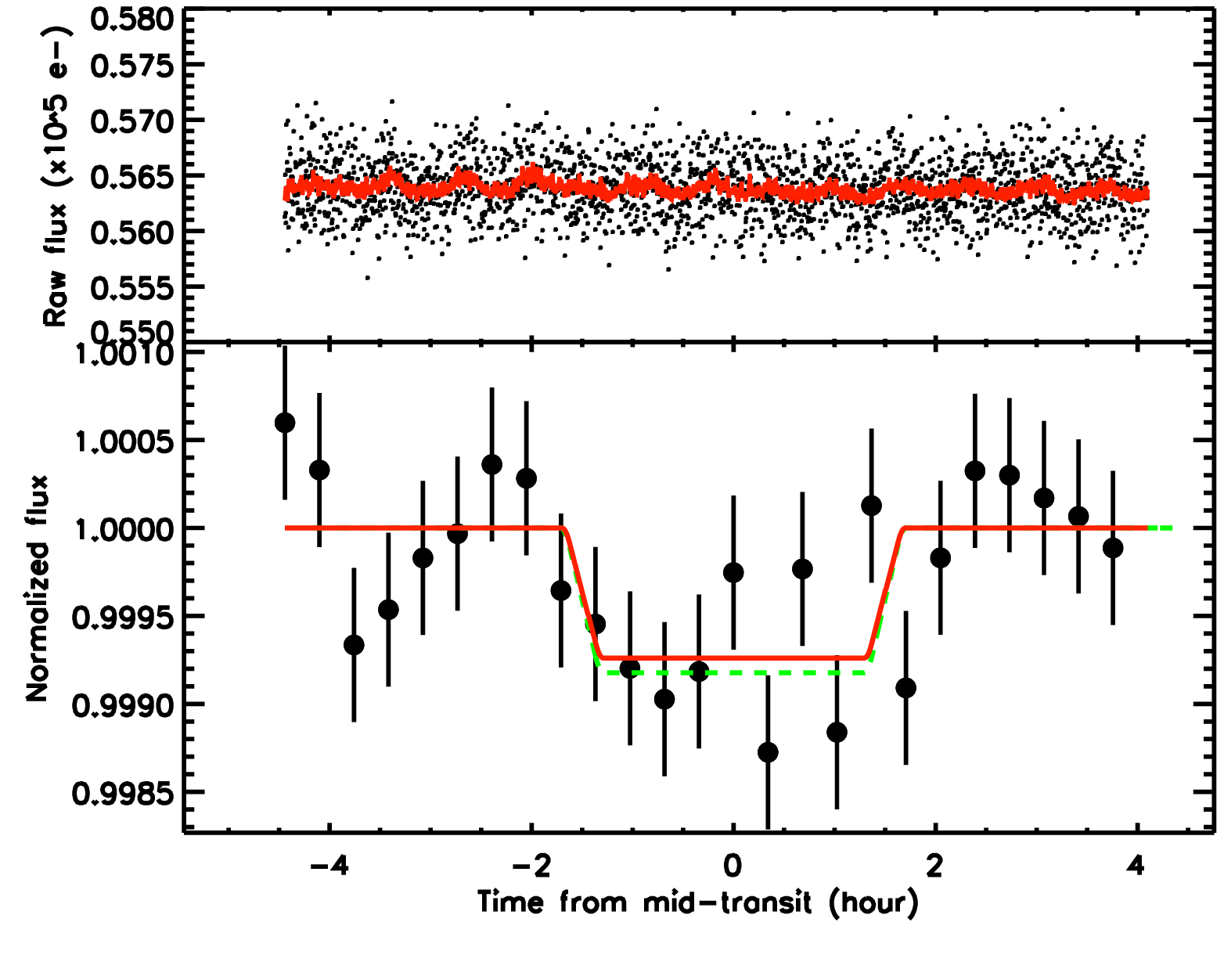}
 \includegraphics[width=4in]{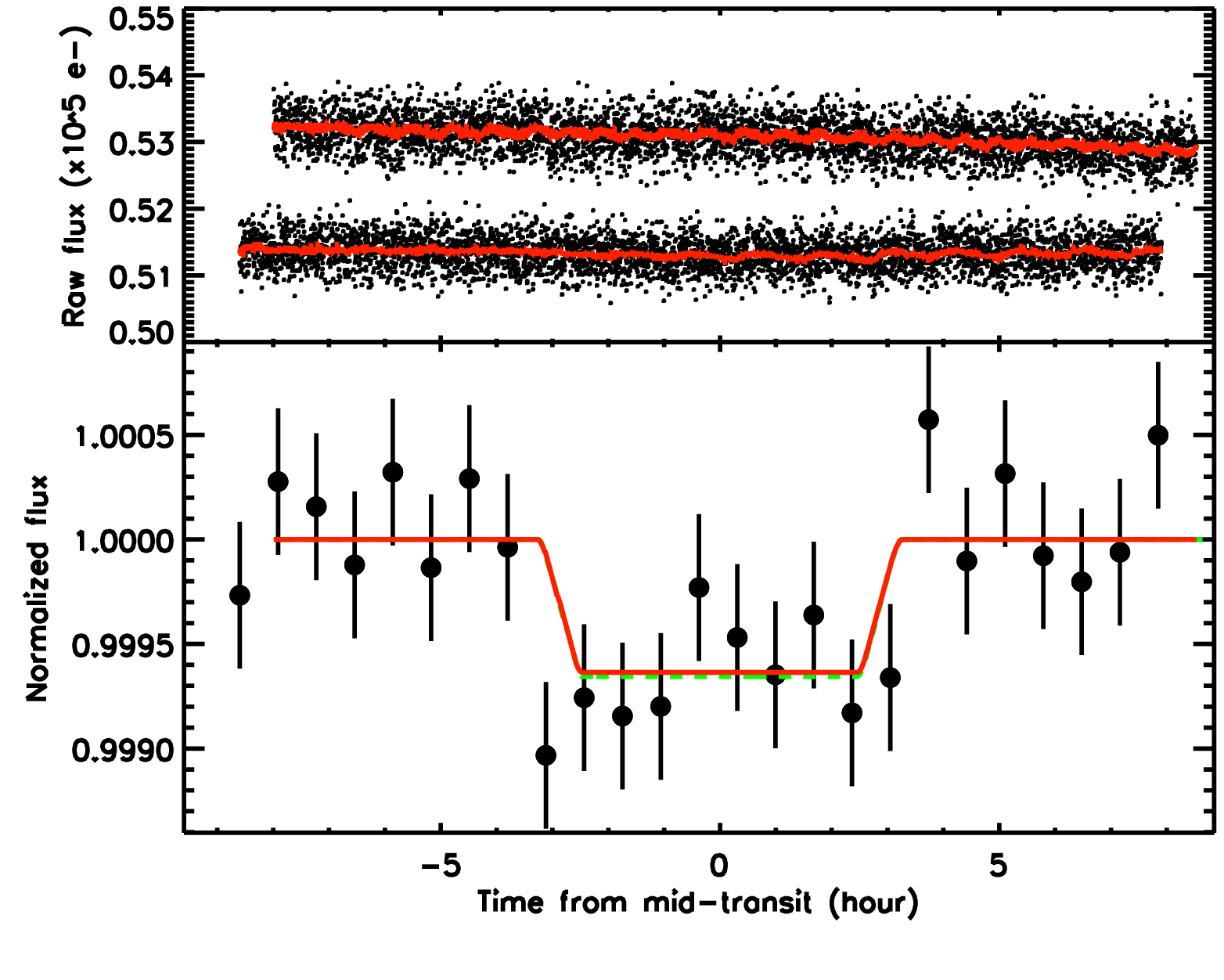}
 \caption{\wspitzer\ transit light curves of \starname\ observed in the IRAC band-pass at 4.5~\micron\ spanning times of transit of \planetc\ ({\it upper half of figure}) and \planetd\ ({\it lower half of figure}). For each candidate, the raw and unbinned time series are shown in the upper panels, and the red solid lines correspond to the best-fit models, which include both the effects of the instrumental variation with time and image position and the planetary transit (see \S\ref{sec:spitzer}). In the lower panel for each candidate, we display these data after correcting
for the instrumental model, normalizing, and binning by 20~minutes. In the case of \planetd\ we co-added the two light curves.  The best-fit model for the transits are plotted in red and the transit shapes expected from the parameters estimated from the \kepler\ observations (Table~\ref{tab:p-params}) are over-plotted as dashed green lines. The transit depths measured in the \spitzer\ and \kepler\ bandpasses agree to better than $1~\sigma$.}
   \label{fig:spitzerlc}
\end{center}
\end{figure}

To obtain an estimate of the correlated and systematic errors in our measurements, we use the residual permutation bootstrap method as described in \cite{desert09}. 
In this method, the residuals of the initial fit are shifted systematically and sequentially by one frame and added to the transit light curve model, which
is then fit once again and the process is repeated. We assign the error bars to be the region containing $34\%$ of the results above and $34\%$ of the results below the median 
of the distributions, as described in \citet{desert11b}.  As we observed two transits of \planetd\, we further evaluated the weighted
mean of the transit depth for this candidate. In Table~\ref{tab:spitzer}, we provide a summary of the \spitzer\ observations and report our estimates of the transit depths and uncertainties.
In Figure~\ref{fig:spitzerlc}, we plot both the raw and corrected time series for each candidate, and overplot the theoretical curve expected
using the parameters estimated from the \kepler\ photometry (see below).

The adaptive optics images described in \S\ref{sec:ao} reveal the
presence of a star adjacent to \starname.  This adjacent star is 4.5
magnitudes fainter in $J$-band that \starname,
and located at an angular separation of $3.8\arcsec$, which corresponds to 3.1 IRAC pixels. 
We tested whether the measured transit depths have to be corrected to take into account the contribution from this stellar companion.
We computed the theoretical dilution factor by extrapolating the $J$-band measurements to the \spitzer\ bandpass at 4.5~\micron. 
We estimate that 1.6\% of the photons recorded during the observation come from the companion star.
For a blend of two sources, the polluted transit depth would be $d/(1+\epsilon)$, where $d$ is the unblended transit depth,
and $\epsilon=1.6$\%. Since the effect on $d$ is well below our detection threshold, we 
conclude that the presence of the contaminant star near \starname\ does not affect our estimates of the transit depths.
 
We calculate the transit shapes that would be expected from the parameters estimated from the \kepler\ photometry (Table~\ref{tab:p-params}) and overplot these
on the \spitzer\ time series in Figure~\ref{fig:spitzerlc}.  The depths we measure with \spitzer\ are in agreement with the depths expected from the \kepler-derived 
parameters at the $1~\sigma$ level. 
Our \spitzer\ observations demonstrate that the transit signals of \planetc\ and \planetd\ are achromatic, as expected for
planetary companions and in conflict with the expectation for most (but not all) astrophysical false positives resulting from
blends of stars within the photometric aperture of \kepler.


\begin{center}
\begin{deluxetable}{lccccc}
\tabletypesize{\scriptsize}
\tablecaption{Transit Depths at 4.5~\micron\ from {\it Warm Spitzer}}
\tablewidth{0pt}
\tablehead{\colhead{Candidate} & \colhead{AOR Name} & \colhead{Date of Observation [UT]} & \colhead{Data Number} & \colhead{Time of Transit Center [BJD]} & \colhead{Transit Depth(\%)} }
\startdata
\planetc & r41165824        & 2010-12-05   & 2291 & 2455536.0209 & $0.075\pm0.015$  \\
\planetd & r39437568        & 2010-09-24   & 4451 & 2455463.4022 & $0.063^{+0.019}_{-0.014}$  \\
\planetd & r41164544        & 2010-12-10   & 4383 & 2455540.9925 & $0.067\pm0.016$  \\   
\planetd & weighted mean & $-$                & $-$    & $-$                   & $0.065\pm0.011$ \\
\enddata
\label{tab:spitzer}
\end{deluxetable}
\end{center}

\clearpage

\subsection{Spectroscopy}\label{sec:spectroscopy}

We obtained 30 high resolution spectra of \starname\ between UT 2009 August 30 and 2011 June 16 
using the HIRES spectrometer on the Keck~I 10-m telescope \citep{vogt94}.  We took spectra with the 
same spectrometer set up of HIRES, and with the same spectroscopic analysis, that we
normally use for precise Doppler work of nearby FGK stars \citep{johnson11}, which 
typically yields a Doppler precision of 1.5~\ms\ for slowly rotating FGKM stars.  Typical exposure
times ranged from 30$-$45 minutes, yielding an SNR of 120 per pixel (1.3~\kms).  
The first 9 observations were made with the B5 decker (0\farcs87 x 3\farcs0) that does not
permit moonlight subtraction.  The remaining 21 observations were made with the C2
decker (0\farcs87 x 14\farcs0) that permits sky subtraction.  The internal errors
were estimated to be between 1.5$-$2~\ms.  
We augmented these uncertainties by adding a jitter term 
of 2.0~\ms\ in quadrature.  The earlier 9 observations are
vulnerable to modest contamination from moonlight, and we have further augmented
the uncertainties for these 9 values by adding in quadrature a term of 2.7~\ms, which
is based on the ensemble performance of stars similarly affected for this magnitude.
The final uncertainties range from 2.5$-$4~\ms.  The estimated RVs and uncertainties are given
in Table~\ref{tab:rvtab}.  We also undertook a study of these spectra to determine the spectral
line bisectors with the goal of placing limits on these sufficient to preclude astrophysical false positives.  
However, we found that the scatter in the bisector centers was somewhat larger than the RV variations,
rendering the RV detection, while sufficient for mass constraint,  inconclusive for confirmation.
We therefore undertook the statistical study described in \S4).

We obtained two additional spectra of \starname\ with HIRES on UT 2009 September 08 and 2010 August 24.
These observations were gathered without the iodine cell, so that the extracted stellar spectrum could be used for the template for the 
RV analysis and for estimating stellar parameters.  The exposure time for the first spectrum was 30 minutes yielding a typical SNR of $90$ and the B1
decker ($R=60,000$) was used.  The observing conditions were slightly better when the second spectrum was
obtained and a one hour exposure yielded a SNR of $140$. The taller B3 decker ($R=60,000$) was used to carry
out a better sky subtraction.  We analyzed these two spectra using the LTE spectral synthesis analysis software \textit{Spectroscopy Made Easy} 
\cite[SME;][]{valenti96, valenti05} to estimate the values of \teff, \logg\, \feh\ and \vsini.  We found that the estimates from each spectrum were
consistent to within $1~\sigma$, and hence we averaged our two estimates to obtain 
$\teff = 5455 \pm 44$~K, $\logg = 4.4 \pm 0.1$, \feh$= 0.01 \pm 0.04$, and $\vsini < 2~\kms$; the errors listed are those resulting from
the analysis of each individual spectrum, and we have refrained from assuming a decrease by a factor of $\sqrt{2}$.
We also proceeded to measure the flux in the cores of the Ca~II~H and K lines to evaluate 
the chromospheric activity.  We measured that the ratio of emission in these lines to the bolometric emission was \logrhk$=-4.93\pm0.05$. 
This estimate was derived from a Mt.\ Wilson-style S-value of $0.183\pm0.005$ \citep{isaacson10}, using the measured color $B-V = 0.725$.  The \logrhk\ value suggests a 
low activity level for a star of this spectral type, which is consistent with the measured $\vsini < 2\ \kms$. Using the relations of \citet{noyes84}
and \citet{mamajek08}, we infer a rotation period of 31~days.

We also gathered three moderate signal-to-noise ratio, high-resolution spectra of
\starname\ for reconnaissance purposes, two with the
FIbre--fed \'Echelle Spectrograph (FIES) at the 2.5\,m Nordic Optical
Telescope (NOT) at La Palma, Spain \citep{djupvik10} and one with the Tull
Coud{\'e} Spectrograph on the McDonald observatory 2.7m  Harlan Smith Telescope.
The FIES spectra were taken on 2009 August 5 and 6 using the medium and high
resolution fibers resulting in a resolution of 46,000 and 67,000,
respectively.  Each spectrum has a wavelength coverage of approximately $360-740$~nm. The
McDonald spectrum was taken on 2010 October 25, with a spectral resolution
of 60,000. This spectrum was exposed to a SNR of $55$ per resolution element for
the specific purpose of deriving reliable atmospheric parameters for the
star. 

As an independent check on the parameters derived from the SME analysis of the Keck/HIRES data described above, 
we derived stellar parameters following \citet{torres02} and \citet{buchhave10}.  As part of this analysis, we employed 
a new fitting scheme that is currently under development by L.~Buchhave, 
allowing us to extract precise stellar parameters from the spectra. We analyzed the two FIES spectra, the McDonald
spectrum and the three HIRES template spectra. These results were found to be consistent within the errors.
Taking the average of the stellar parameters from the different instruments yielded the following parameter estimates:
\teff $= 5563 \pm 50$~K, \logg $=4.52 \pm 0.10$, [m/H]$=+0.04 \pm 0.08$, and \vsini $=1.80 \pm 0.50~\kms$, 
in agreement with the parameters from SME within the uncertainties. The average systemic radial 
velocity of the six observations was $-21.87 \pm 0.96~\kms$ on the IAU standard scale, which includes the correction
for the gravitational redshift of the Sun.

We note that the two analyses yielded consistent results for \logg, metallicity, and \vsini, but that the estimates of \teff\
differed by twice the formal error. Hence we elected to adopt the results of the SME analysis for our final values,
but we increased the uncertainty on \teff\ to 100~K to reflect the difference between the two estimates.  We list our estimates for the spectroscopically determined
parameters in Table~\ref{tab:s-params}.

\begin{center}
\begin{deluxetable}{lcc}
\tabletypesize{\scriptsize}
\tablecaption{Keck HIRES Radial Velocity Measurements for \starname}
\tablewidth{0pt}
\tablehead{\colhead{Date of Obs. [BJD]} & \colhead{Radial Velocity [\ms]} & \colhead{Uncertainty [\ms]\tablenotemark{a}}}
\startdata
2455073.885713 &      $-$5.86  &     3.78 \\
2455074.861139 &      $-$2.94  &     3.74 \\
2455075.906678 &       1.09   &    3.75 \\
2455076.883792 &       2.57   &    3.84 \\
2455077.907884 &      11.08   &    3.82 \\
2455081.952637 &      $-$1.51  &     3.83 \\
2455082.832587 &      $-$9.53  &     3.78 \\
2455083.888144 &      $-$7.17  &     3.84 \\
2455084.893633 &       4.59    &   3.88 \\
2455134.766132 &     $-$14.15  &     3.88 \\
2455314.027873 &       2.05    &   2.54 \\
2455320.085390 &      $-$9.47  &     2.59 \\
2455321.072318 &      $-$4.55  &     2.56 \\
2455345.011208 &      $-$8.71  &     2.82 \\
2455351.073218 &       0.41    &   2.70 \\
2455352.043110 &      $-$1.84  &     2.50 \\
2455372.898433 &       1.71    &   2.62 \\
2455374.967141 &     $-$16.20  &     2.62 \\
2455378.007402 &      $-$2.59  &     2.53 \\
2455380.950103 &      $-$0.62  &     2.55 \\
2455403.055761 &       7.59    &   2.96 \\
2455407.903131 &     $-$10.15  &     2.64 \\
2455411.959424 &     $-$13.79  &     2.65 \\
2455413.894286 &      $-$2.60  &     2.59 \\
2455433.818795 &       4.81    &   2.46 \\
2455435.841063 &      10.04    &   2.58 \\
2455440.792117 &     $-$10.94  &     2.50 \\
2455500.817440 &      $-$2.94  &     3.34 \\
2455522.725846 &      $-$6.56  &     2.63 \\
2455724.055721 &      $-$2.35  &     2.64 \\                                 
\enddata
\tablenotetext{a}{Includes jitter of 2 \ms}
\label{tab:rvtab}
\end{deluxetable}
\end{center}

\clearpage

\section{Validation of the Planets \planetb, \planetc, and \planetd}
\label{sec:blender}

While the analysis of the radial-velocity (\S\ref{sec:spectroscopy}) 
data yielded detections for \planetb\ and \planetc, we found that 
our analyses of the bisector spans were not sufficient to confirm the planetary origin of
those variations.  Moreover, for \planetd, \koie, and \koif\ there is
no Doppler detection. We therefore rely on a fundamentally different technique to
establish which, if any, of these signals can be persuasively attributed to planets.  
As explained in \cite{lissauer11b}, when dynamical confirmation of a
planet candidate by the radial velocity method or by transit timing variations cannot be
achieved, we attempt instead to validate the candidate by tabulating
all viable false positives (blends) that could mimic the signal. We then assess the 
likelihood of these blends, and compare it with an {\it
a priori\/} estimate of the likelihood that the signal is due to a
true planet. We consider the signal to be validated when the likelihood of a planet exceeds that of a false positive
by a sufficiently large ratio, typically at least 300 (i.e. $3~\sigma$).

Our tabulation of the viable scenarios resulting from blends was accomplished 
with the \blender\ algorithm \citep{torres04, torres11, fressin11a, fressin11b}
combined with some of the follow-up observations described earlier
(high-resolution imaging, centroid motion analysis, spectroscopy, and
\spitzer\ observations). \blender\ attempts to fit the \kepler\
photometry with a vast array of synthetic light curves generated from
blend configurations consisting of chance alignments with background
or foreground eclipsing binaries (EBs), as well as eclipsing binaries
physically associated with \starname\ (hierarchical triples). We also
considered cases in which the second star is eclipsed by a larger
planet, rather than by another star. A wide range of parameters is
explored for the eclipsing pair, as well as for the relative distance
separating it from the target.  Scenarios giving poor fits to the data
(specifically, a $\chi^2$ value that indicates a discrepancy of at least $3~\sigma$ worse 
than that corresponding to the transiting planet model) are considered to be ruled out.  For full details 
of this technique we refer the reader to the above sources.

The combination of the shorter periods and deeper transits for \planetb\ and
\planetc\ results in higher SNR for those signals
compared to the others. Consequently the shape of the transit is
better defined, and this information makes it easier to reject false
positives with \blender, as we show below. The transit depths of \koie\ and \koif\
are only 82 and 101 parts per million; this renders these signals far more
challenging to validate, and we find below that we are currently not able to demonstrate
unambiguously that these two signals are planetary in origin.  \planetd\ is similar in depth
to \planetb, but due to its longer orbital period, far fewer transits have been
observed.  This results in a lower SNR in the phase light curve. 
We begin by describing this case.

\begin{figure}
\begin{center}
\includegraphics[width=2.6in]{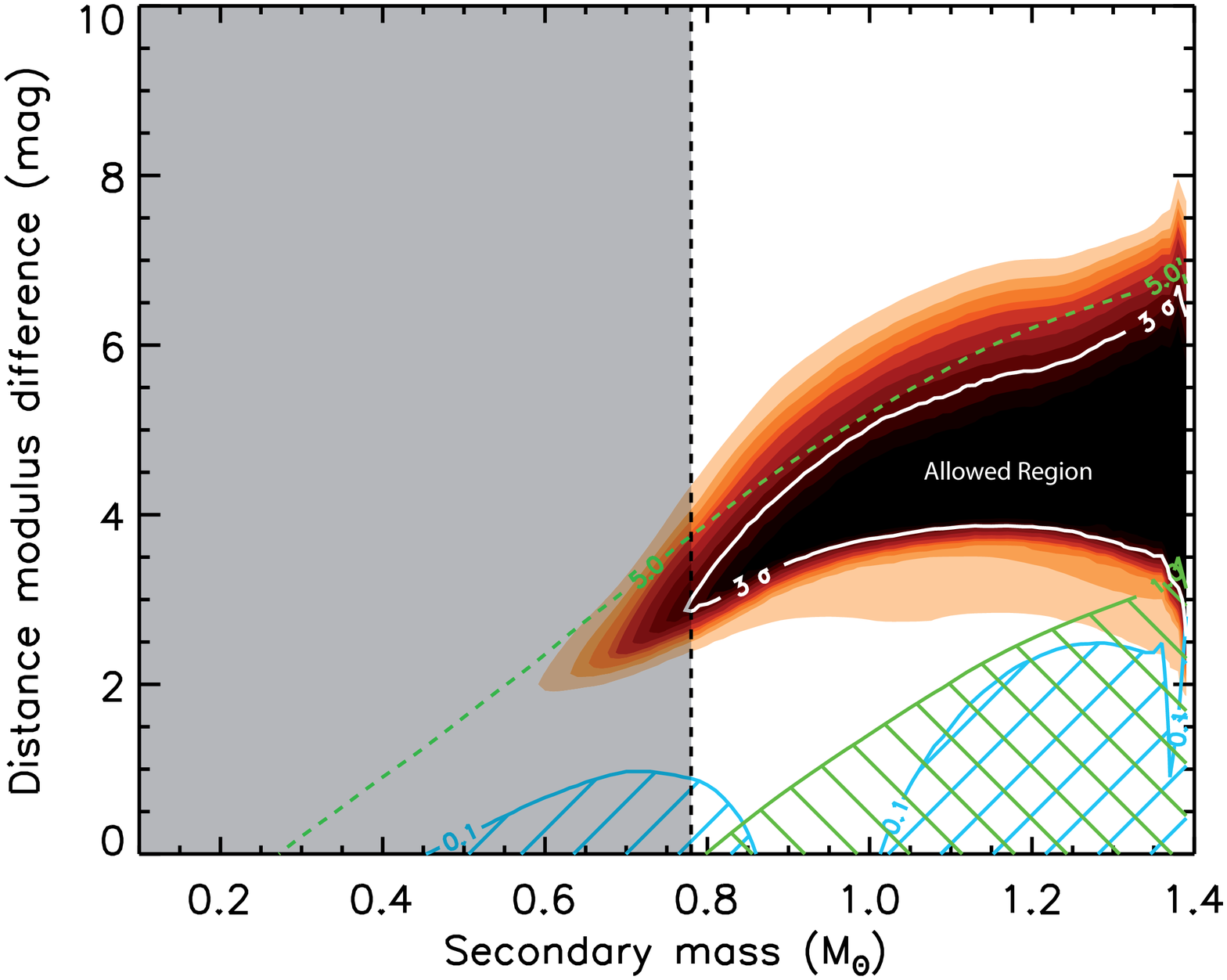}
\includegraphics[width=2.6in]{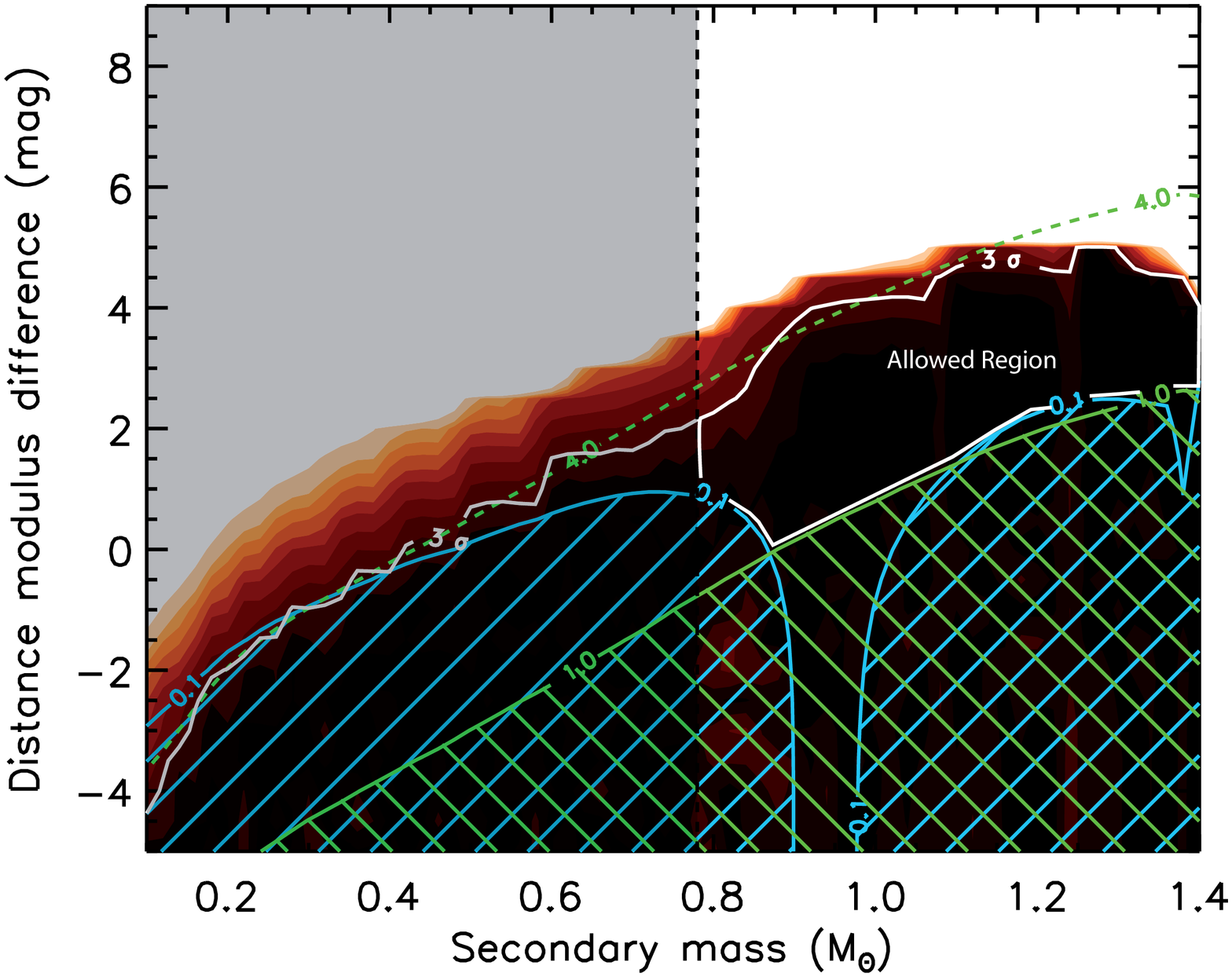}
\includegraphics[width=2.6in]{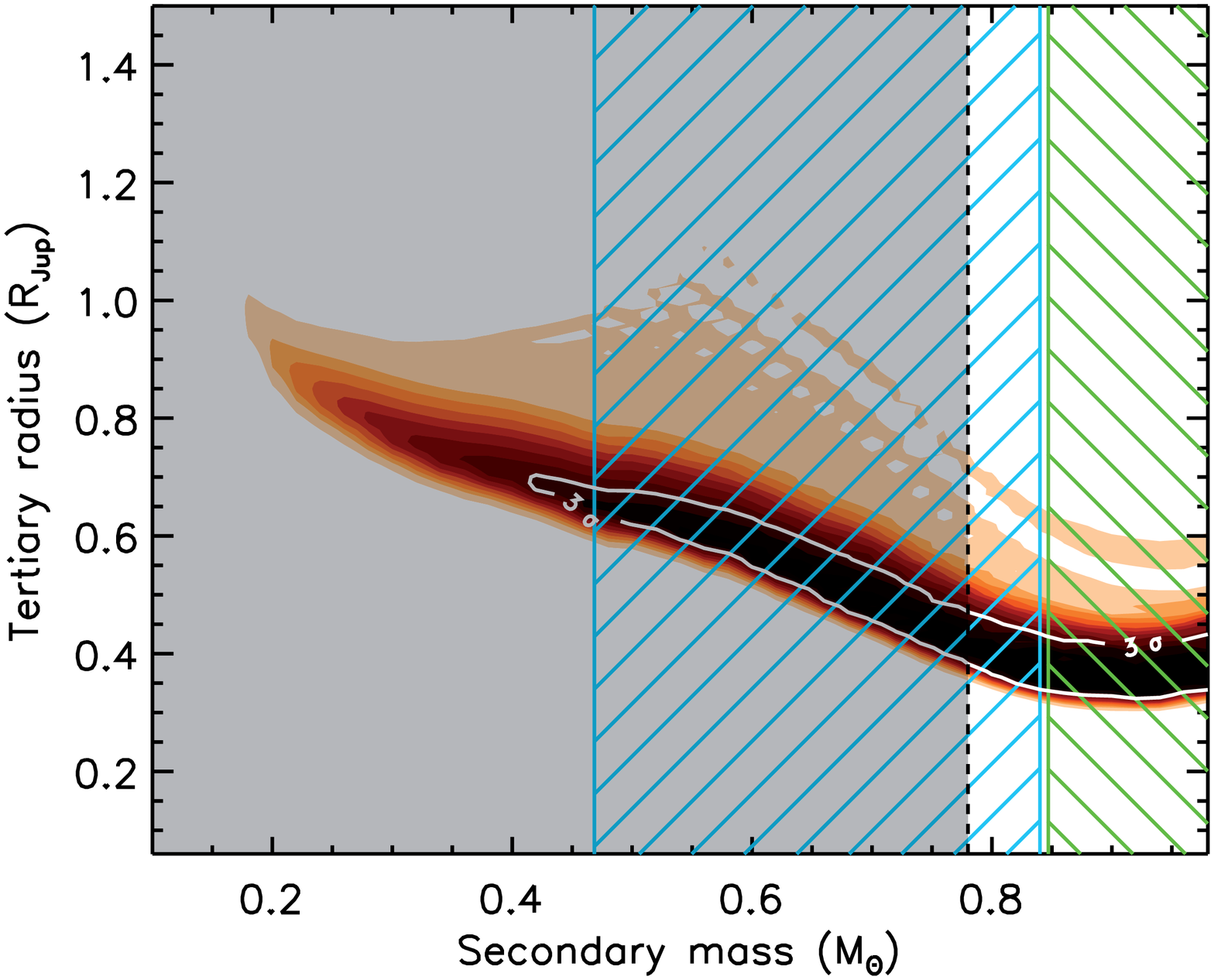}
\caption{\blender\ goodness-of-fit contours corresponding to
three different blend scenarios for \planetd: background EBs
(\emph{top left}), background or foreground stars transited by a larger
planet (\emph{top right}), and stars physically associated with the
target that are transited by a larger planet (\emph{bottom}). The mass
of the intruding star (referred to in the \blender\ nomenclature as
the secondary) is shown along the horizontal axis, and the distance
between this star and the target is shown on the vertical axis of the
top two panels, expressed for convenience in terms of the distance
modulus difference. The vertical axis in the bottom panel shows the
sizes of the planets (tertiaries) orbiting physically associated
stars.  Viable blend models are those giving fits with $\chi^2$ values
within $3~\sigma$ of the best planet fit, and lie inside the white
contours. Other colored areas outside the white contours indicate
regions of parameter space with increasingly worse fits to the data. Blends excluded by our \spitzer\
constraints are shown with the shaded gray area (secondary masses $<
0.78\,M_{\sun}$). Blue cross-hatched areas indicate regions in which
blends are excluded because they are either too red (left) or too blue
(right) compared to the measured $r-K_s$ color of \starname. Blend
scenarios in the green cross-hatched areas are also ruled out because
they are within $\Delta Kp = 1.0$ of the brightness of the target,
and would have been detected spectroscopically. The dashed diagonal
green lines in the top panels mark the faintest blends that give
acceptable fits to the light curve, corresponding to $\Delta Kp
\approx 5$ in the top left panel and $\Delta Kp \approx 4$ in the top right
panel. As a result of the combined constraints from our \spitzer\
observations, color index, and brightness, \emph{all} physically
associated triples are excluded.
\label{fig:blender_70.03}}
\end{center}
\end{figure}

Figure~\ref{fig:blender_70.03} illustrates the \blender\ results for \planetd. 
The three panels represent cuts through the space of
parameters for blends consisting of background EBs,
background or foreground stars transited by larger planets, and
physically associated triples. In the latter case we find that the
only scenarios able to mimic the signal are those in which the
companion star is orbited by a larger planet, rather than another
star. The orange-red-brown-black shaded regions correspond to
different levels of the $\chi^2$ difference between blend models and
the best transiting planet fit to the \kepler\ data, expressed in
terms of the statistical significance of the difference ($\sigma$).
The $3~\sigma$ level is represented by the white contour, and only
blends inside it ($<~3\sigma$) are considered to give acceptable fits
to the \kepler\ photometry.  Other constraints further restrict the
area allowed for blends.  The green hatched areas are excluded because
the EB is within one magnitude of the target in the $Kp$ band, and
would generally have been noticed in our spectroscopic
observations. The blue hatched areas are also excluded because the
overall color of the blend is either too red (left in the top two
panels) or too blue (right) compared to the measured Sloan-2MASS
$r-K_s$ color of \starname, as listed in the \kepler\ Input Catalog
\citep[KIC;][]{brown11}. Additionally, \spitzer\ observations rule
out blends involving EBs (or star+planet pairs) with stars less
massive than about 0.78\,$M_{\sun}$ (gray shaded area to the left of
the vertical dotted line), because the predicted depth of the transits
in the 4.5\,\micron\ bandpass of \wspitzer\ would be more than
$3~\sigma$ larger than our \spitzer\ observations indicate. Note that
the combination of these constraints rules out \emph{all} physically
associated triple configurations for \planetd, so that only certain blend
scenarios involving background EBs or background/foreground stars
transited by larger planets present suitable alternatives to a true
planet model.

We estimate the frequency of these remaining blends following
\cite{torres11} and \cite{fressin11a}, as the product of three
factors: the expected number density of stars in the vicinity of
\starname, the area around the target within which we would miss such
stars, and an estimate of how often we expect those stars to be in EBs
or be transited by a larger planet of the right characteristics (specified
by the stellar masses, planetary sizes, orbital eccentricities, and other
characteristics as tabulated by \blender). For the number densities we appeal to the
Besan\c{c}on Galactic structure model of \cite{robin03}. Constraints
from our high-resolution imaging (see \S\ref{sec:ao}) allow us to estimate
the maximum angular separation ($\rho_{\rm max}$) at which blended
stars would be undetected, as a function of brightness. We derive our estimates of
the frequencies of EBs and larger transiting planets involved in
blends from recent studies by the \kepler\ Team
\citep{slawson11, borucki11}, in the same way as done for our
earlier studies of Kepler-9\,d, Kepler-10\,c, and Kepler-11\,g
\cite[see][]{torres11, fressin11a, lissauer11b}.

The results of our calculations for \planetd, performed in half-magnitude
bins, are shown in Table~\ref{tab:blendfreq_70.03} separately for
background EBs and for background or foreground stars transited by a
larger planet. The first two columns give the $Kp$ magnitude range
of each bin, and the magnitude difference $\Delta Kp$ relative to
the target, calculated at the faint end of each bin. Column~3 reports
the stellar density near the target, subject to the mass constraints
from \blender\ as shown in Figure~\ref{fig:blender_70.03}. Column~4
gives the maximum angular separation at which background stars would
escape detection in our imaging observations. In this particular case
those observations are more constraining than the $3~\sigma$ exclusion
limit set by our analysis of the flux centroids (0\farcs65; see \S\ref{sec:ao}). 
The product of the area implied by
$\rho_{\rm max}$ and the densities in the previous column are listed
in column~5, in units of $10^{-6}$. Column~6 is the result of
multiplying this number of stars by the frequency of suitable EBs
\citep[$f_{\rm EB} = 0.78$\%; see][]{fressin11a}.  A similar
calculation is performed for blends involving stars transited by
larger planets, and is presented in columns~7--10, using $f_{\rm
planet} = 0.18\%$. The latter is the frequency of planets in the
radius range allowed by \blender\ for these types of scenarios, which
is 0.4--2.0\,$R_{\rm Jup}$ \citep[see][]{borucki11}.  The sum of the
contributions in each bin is given at the bottom of columns~6 and
10. The total number of blends (i.e., the blend frequency) we expect
{\it a priori} is reported in the last line of the table by adding
these two numbers together, and is approximately ${\rm BF} = 6.0
\times 10^{-7}$.

We now compare this estimate with the likelihood that \planetd\ is a true
transiting planet (planet prior).  To calculate the planet prior we
again make use of the census of 1,235 candidates reported by
\cite{borucki11} among the 156,453 \kepler\ targets observed during
the first four months of operation of the Mission\footnote{While
these 1,235 candidates have not yet been confirmed as true planets,
the rate of false positives is expected to be quite low
\citep[10\% or less; see][]{morton11}, so our results will not
be significantly affected by the assumption that all of the candidates
are planets. We further assume here that the census of
\cite{borucki11} is complete at these planetary radii.}.  We count 100 candidates that are within
$3~\sigma$ of the measured radius ratio of \planetd, implying an {\it a
priori} transiting planet frequency of ${\rm PF} = 100/156,\!453 = 6.4
\times 10^{-4}$.  The likelihood of a planet is therefore several
orders of magnitude larger than the likelihood of a false positive
(${\rm PF/BF} = 6.4 \times 10^{-4}/6.0 \times 10^{-7} \approx 1100$),
and we consider \planetd\ to be validated as a planet with a high degree
of confidence.

\begin{figure}
\includegraphics[width=6in]{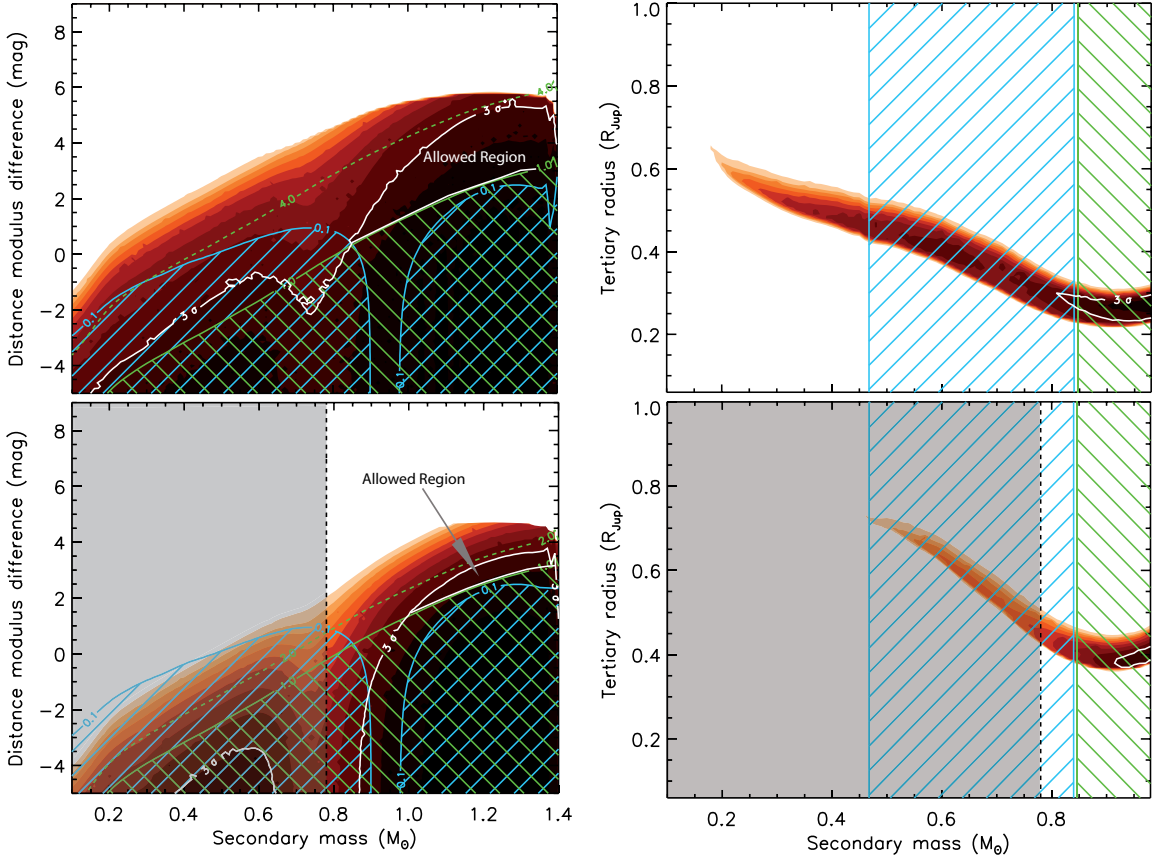}
\caption{\blender\ constraints for
\planetb\ {\it(top panels)} and \planetc\ {\it (bottom)}, showing chance alignments
with a star+planet pair on the left and physically associated stellar
companions transited by a larger planet on the right. See
Figure~\ref{fig:blender_70.03} for the meaning of the various
lines. The space of parameters for background EBs is not shown as all
of those scenarios provide very poor fits to the transit light curve,
and are ruled out. All blends involving physically associated stars
transited by a larger planet {\it (right panels)} are excluded by
a combination of spectroscopic constraints (specifically, on the absence of
a secondary spectrum) and color constraints.\label{fig:blender_70.02_70.01}}
\end{figure}

The transit signals from \planetb\ and \planetc\ are better defined, and as a result
\blender\ is able to rule out all scenarios involving background EBs
consisting of two stars, as well as all physically associated triples.
This reduces the blend frequencies by one to two orders of magnitude
compared to \planetd. For \planetc, \spitzer\ observations are available as
well, although the constraints they provide are redundant with color
information also available for the star, which already rules out
contaminants of late spectral type. The areas of parameter space in
which \blender\ finds false positives providing acceptable fits to the
photometry are shown in Figure~\ref{fig:blender_70.02_70.01}.  The
detailed calculations of the blend frequencies for \planetb\ and \planetc\ are
presented in Table~\ref{tab:blendfreq_70.02} and
Table~\ref{tab:blendfreq_70.01}, respectively, using appropriate
ranges for the larger planets orbiting the blended stars as allowed by
\blender, along with the corresponding transiting planet frequencies
specified in the headings of column~6.

Planet priors for these two candidates were computed as before using
the catalog of \cite{borucki11}. We count 52 cases in that list
within $3~\sigma$ of the measured radius ratio of \planetb, leading to an
\emph{a priori} planet frequency of $52/156,\!453 = 3.3 \times
10^{-4}$.  This is nearly 20,000 times larger than the blend frequency
given in Table~\ref{tab:blendfreq_70.02} (${\rm BF} = 1.7 \times
10^{-8}$). For \planetc\ the planet prior based on the measured radius
ratio is
$28/156,\!453 = 1.8 \times 10^{-4}$, which is approximately $10^5$ times larger
than the likelihood of a blend. Therefore, both \planetb\ and \planetc\ are
validated as planets with a very high degree of confidence.

We carried out similar calculations for the candidates \koie\ and
\koif.  The transit signals of these two candidates are much more
shallow than those of \planetb, \planetc, and \planetd.
As a result, 
the constraint on the shape of the transit is
considerably weaker than in the cases described above, and many more
false positives than before are found with \blender\ that provide
acceptable fits within $3~\sigma$ of the quality of a planet
model. Additionally, neither of these candidates were observed with
\spitzer, so the constraint on the near-infrared depth of the transit
that allowed us to rule out some of the blends for \planetd\ is not
available here. In particular, physically associated stars transited
by a larger planet cannot all be ruled out, and this ends up
contributing significantly to the overall blend frequency. We conclude that
the \blender\ methodology as implemented above is insufficient to 
validate either \koie\ or \koif, and we defer this issue to a subsequent
study \citep{fressin12}.

\begin{landscape}
\begin{deluxetable}{ccccccccccc}
\tabletypesize{\scriptsize}
\tablewidth{0pc}
\tablecaption{Blend frequency estimate for \planetd\ (\koid).\label{tab:blendfreq_70.03}}
\tablehead{
  & & \multicolumn{4}{c}{Blends Involving Stellar Tertiaries}
  & & \multicolumn{4}{c}{Blends Involving Planetary Tertiaries} \\ [+1.5ex]
\cline{3-6} \cline{8-11} \\ [-1.5ex]
\colhead{$Kp$ Range} &
\colhead{$\Delta Kp$} &
\colhead{Stellar Density\tablenotemark{a}} &
\colhead{$\rho_{\rm max}$} &
\colhead{Stars} &
\colhead{EBs} & &
\colhead{Stellar Density\tablenotemark{a}} &
\colhead{$\rho_{\rm max}$} &
\colhead{Stars} &
\colhead{Blends ($\times 10^{-6}$)}
\\
\colhead{(mag)} &
\colhead{(mag)} &
\colhead{(per sq.\ deg)} &
\colhead{(\arcsec)} &
\colhead{($\times 10^{-6}$)} &
\colhead{$f_{\rm EB} = 0.78$\%} & &
\colhead{(per sq.\ deg)} &
\colhead{(\arcsec)} &
\colhead{($\times 10^{-6}$)} &
\colhead{$R_p \in \left[0.4\!-\!2.0\,R_{\rm Jup}\right]$, $f_{\rm Plan}=0.18$\%}
\\
\colhead{(1)} &
\colhead{(2)} &
\colhead{(3)} &
\colhead{(4)} &
\colhead{(5)} &
\colhead{(6)} & &
\colhead{(7)} &
\colhead{(8)} &
\colhead{(9)} &
\colhead{(10)}
}
\startdata
12.5--13.0  &  0.5 &\nodata &\nodata  & \nodata & \nodata && \nodata&\nodata & \nodata& \nodata \\
13.0--13.5  &  1.0 &\nodata &\nodata  & \nodata & \nodata && \nodata&\nodata & \nodata& \nodata \\
13.5--14.0  &  1.5 &  28    & 0.075   & 0.038   & 0.0003  &&  273   & 0.075  &  0.372 & 0.0007  \\
14.0--14.5  &  2.0 &  77    & 0.093   & 0.164   & 0.0013  &&  423   & 0.093  &  0.901 & 0.0016  \\
14.5--15.0  &  2.5 & 119    & 0.11    & 0.349   & 0.0028  &&  572   & 0.11   &  1.678 & 0.0024  \\
15.0--15.5  &  3.0 & 238    & 0.13    & 0.975   & 0.0077  &&  897   & 0.13   &  3.675 & 0.0053  \\
15.5--16.0  &  3.5 & 532    & 0.15    & 2.902   & 0.0229  &&  1183  & 0.15   &  6.452 & 0.0093  \\
16.0--16.5  &  4.0 & 1321   & 0.20    & 12.81   & 0.1012  &&  1675  & 0.20   &  16.24 & 0.0234  \\
16.5--17.0  &  4.5 & 1593   & 0.25    & 24.13   & 0.1907  && \nodata&\nodata & \nodata& \nodata \\
17.0--17.5  &  5.0 & 1295   & 0.30    & 28.25   & 0.2232  && \nodata&\nodata & \nodata& \nodata \\
17.5--18.0  &  5.5 &\nodata &\nodata  & \nodata & \nodata && \nodata&\nodata & \nodata& \nodata \\
18.5--19.0  &  6.0 &\nodata &\nodata  & \nodata & \nodata && \nodata&\nodata & \nodata& \nodata \\
19.0--19.5  &  6.5 &\nodata &\nodata  & \nodata & \nodata && \nodata&\nodata & \nodata& \nodata \\
19.5--20.0  &  7.0 &\nodata &\nodata  & \nodata & \nodata && \nodata&\nodata & \nodata& \nodata \\
20.0--20.5  &  7.5 &\nodata &\nodata  & \nodata & \nodata && \nodata&\nodata & \nodata& \nodata \\
20.5--21.0  &  8.0 &\nodata &\nodata  & \nodata & \nodata && \nodata&\nodata & \nodata& \nodata \\
\noalign{\vskip 6pt}
\multicolumn{2}{c}{Totals} & 5203 &\nodata &  69.62  & {\bf 0.5511} && 5023  &\nodata & 29.32  & {\bf 0.0526}   \\
\noalign{\vskip 4pt}
\hline
\noalign{\vskip 4pt}
\multicolumn{11}{c}{Blend frequency (BF) = $(0.5511 + 0.0526)\times 10^{-6} \approx 6.04 \times 10^{-7}$} \\ 
\enddata

\tablenotetext{a}{The number densities in Columns 3 and 7 differ
because of the different secondary mass ranges permitted by \blender\
for the two kinds of blend scenarios, as shown in the top two panels of
Figures~\ref{fig:blender_70.03}.}
\tablecomments{Magnitude bins with no entries correspond to brightness 
ranges in
which all blends are ruled out by a combination of \blender\ and other 
constraints.}

\end{deluxetable}
\clearpage
\end{landscape}

\clearpage

\begin{deluxetable}{cccccc}
\tabletypesize{\scriptsize}
\tablewidth{0pc}
\tablecaption{Blend frequency estimate for \planetb\ (\koib).\label{tab:blendfreq_70.02}}
\tablehead{
  & & \multicolumn{4}{c}{Blends Involving Planetary Tertiaries} \\ [+1.5ex]
\cline{3-6} \\ [-1.5ex]
\colhead{$Kp$ Range} &
\colhead{$\Delta Kp$} &
\colhead{Stellar Density} &
\colhead{$\rho_{\rm max}$} &
\colhead{Stars} &
\colhead{Blends ($\times 10^{-6}$)}
\\
\colhead{(mag)} &
\colhead{(mag)} &
\colhead{(per sq.\ deg)} &
\colhead{(\arcsec)} &
\colhead{($\times 10^{-6}$)} &
\colhead{$R_p \in \left[0.27\!-\!1.81\,R_{\rm Jup}\right]$, $f_{\rm Plan}=0.29$\%}
\\
\colhead{(1)} &
\colhead{(2)} &
\colhead{(3)} &
\colhead{(4)} &
\colhead{(5)} &
\colhead{(6)}
}
\startdata
12.5--13.0  &  0.5  & \nodata  &  \nodata &  \nodata &  \nodata \\
13.0--13.5  &  1.0  & \nodata  &  \nodata &  \nodata &  \nodata \\
13.5--14.0  &  1.5  & 271      &  0.075   &  0.370   &  0.00068 \\
14.0--14.5  &  2.0  & 379      &  0.093   &  0.807   &  0.00149 \\
14.5--15.0  &  2.5  & 404      &  0.11    &  1.185   &  0.00218 \\
15.0--15.5  &  3.0  & 464      &  0.13    &  1.901   &  0.00351 \\
15.5--16.0  &  3.5  & 498      &  0.15    &  2.716   &  0.00501 \\
16.0--16.5  &  4.0  & 255      &  0.20    &  2.473   &  0.00456 \\
16.5--17.0  &  4.5  & \nodata  &  \nodata &  \nodata &  \nodata \\
17.0--17.5  &  5.0  & \nodata  &  \nodata &  \nodata &  \nodata \\
17.5--18.0  &  5.5  & \nodata  &  \nodata &  \nodata &  \nodata \\
18.0--18.5  &  6.0  & \nodata  &  \nodata &  \nodata &  \nodata \\
18.5--19.0  &  6.5  & \nodata  &  \nodata &  \nodata &  \nodata \\
19.0--19.5  &  7.0  & \nodata  &  \nodata &  \nodata &  \nodata \\
19.5--20.0  &  7.5  & \nodata  &  \nodata &  \nodata &  \nodata \\
20.0--20.5  &  8.0  & \nodata  &  \nodata &  \nodata &  \nodata \\
\noalign{\vskip 6pt}
\multicolumn{2}{c}{Totals}  & 2669  &\nodata & 9.452 & {\bf 0.0174} \\
\noalign{\vskip 4pt}
\hline
\noalign{\vskip 4pt}
\multicolumn{6}{c}{Blend frequency (BF) = $1.74 \times 10^{-8}$} \\ 
\enddata

\tablecomments{Magnitude bins with no entries correspond to brightness 
ranges in
which all blends are ruled out by a combination of \blender\ and other 
constraints.}

\end{deluxetable}

\clearpage

\begin{deluxetable}{cccccc}
\tabletypesize{\scriptsize}
\tablewidth{0pc}
\tablecaption{Blend frequency estimate for \planetc\ (\koic).\label{tab:blendfreq_70.01}}
\tablehead{
  & & \multicolumn{4}{c}{Blends Involving Planetary Tertiaries} \\ [+1.5ex]
\cline{3-6} \\ [-1.5ex]
\colhead{$Kp$ Range} &
\colhead{$\Delta Kp$} &
\colhead{Stellar Density} &
\colhead{$\rho_{\rm max}$} &
\colhead{Stars} &
\colhead{Blends ($\times 10^{-6}$)}
\\
\colhead{(mag)} &
\colhead{(mag)} &
\colhead{(per sq.\ deg)} &
\colhead{(\arcsec)} &
\colhead{($\times 10^{-6}$)} &
\colhead{$R_p \in \left[0.39\!-\!1.95\,R_{\rm Jup}\right]$, $f_{\rm Plan}=0.18$\%}
\\
\colhead{(1)} &
\colhead{(2)} &
\colhead{(3)} &
\colhead{(4)} &
\colhead{(5)} &
\colhead{(6)}
}
\startdata
12.5--13.0  &  0.5  & \nodata &       \nodata &       \nodata &       \nodata \\
13.0--13.5  &  1.0  & \nodata &       \nodata &       \nodata &       \nodata \\
13.5--14.0  &  1.5  & 221     &       0.075   &       0.300   &       0.00056 \\
14.0--14.5  &  2.0  & 274     &       0.093   &       0.551   &       0.00108 \\
14.5--15.0  &  2.5  & \nodata &       \nodata &       \nodata &       \nodata \\
15.0--15.5  &  3.0  & \nodata &       \nodata &       \nodata &       \nodata \\
15.5--16.0  &  3.5  & \nodata &       \nodata &       \nodata &       \nodata \\
16.0--16.5  &  4.0  & \nodata &       \nodata &       \nodata &       \nodata \\
16.5--17.0  &  4.5  & \nodata &       \nodata &       \nodata &       \nodata \\
17.0--17.5  &  5.0  & \nodata &       \nodata &       \nodata &       \nodata \\
17.5--18.0  &  5.5  & \nodata &       \nodata &       \nodata &       \nodata \\
18.0--18.5  &  6.0  & \nodata &       \nodata &       \nodata &       \nodata \\
18.5--19.0  &  6.5  & \nodata &       \nodata &       \nodata &       \nodata \\
19.0--19.5  &  7.0  & \nodata &       \nodata &       \nodata &       \nodata \\
19.5--20.0  &  7.5  & \nodata &       \nodata &       \nodata &       \nodata \\
\noalign{\vskip 6pt}
\multicolumn{2}{c}{Totals}  & 495  &\nodata & 0.851 & {\bf 0.00164}   \\
\noalign{\vskip 4pt}
\hline
\noalign{\vskip 4pt}
\multicolumn{6}{c}{Blend frequency (BF) = $1.64 \times 10^{-9}$} \\ 
\enddata

\tablecomments{Magnitude bins with no entries correspond to brightness 
ranges in which all blends are ruled out by a combination of \blender\ and other 
constraints.}

\end{deluxetable}

\clearpage

\section{Constraints on Transit Times and Long-term Stability}

In this section we discuss the transit times and long-term stability
of the system of planets orbiting \starname. Both are consistent with the planet interpretation
for all 5 candidates: transit timing
variations \citep[TTVs;][]{holman05, agol05} are not seen or expected, and the system is expected to be 
stable over long timescales.

The individual transit times are measured by allowing a template
transit light curve to slide in time to fit the data for each transit \citep{ford11}.  The resulting
transit times are given in Table \ref{tab:ttv}. Aside from slightly
more scatter than expected from the formal error bars there is no
indication of perturbations such as coherent patterns.  Such excess
scatter is not atypical of transit times measured by the standard
pipeline \citep{ford11}.  Thus, we find no evidence for dynamical
interactions among either the transiting planets or additional,
non-transiting planets.

\begin{figure}
\begin{center}
\includegraphics[scale=0.8]{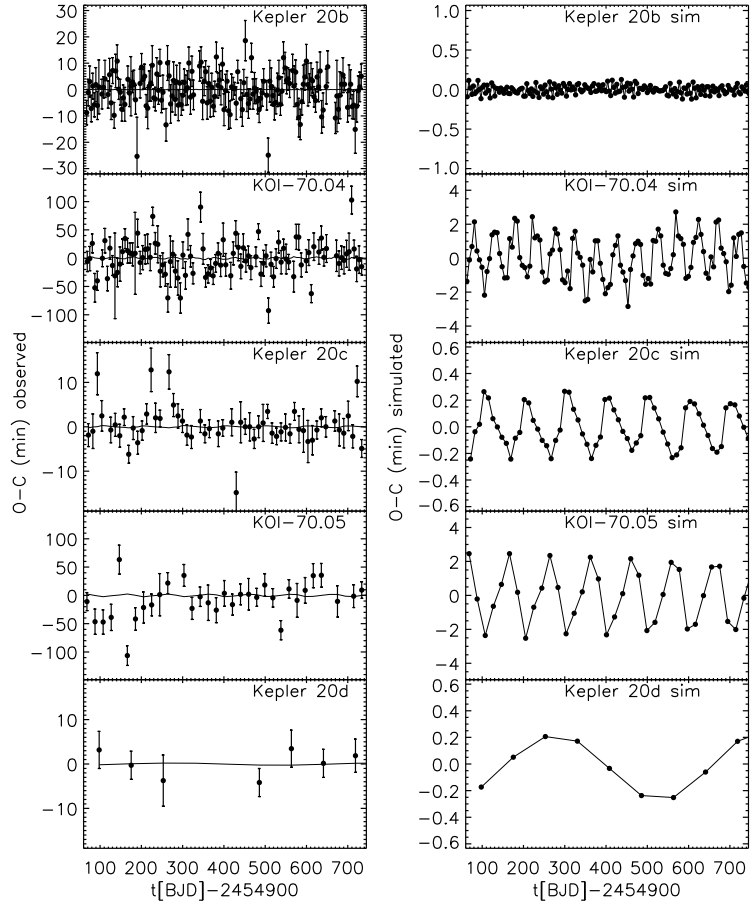}
\caption{Measured and predicted transit timing for the planets of \starname. \emph{Left
panels}: Observed times minus calculated times according to a constant-period model
($O-C$) are plotted as points with error bars, versus transit time. The timing
simulations using circular, coplanar planets with nominal masses are plotted
as lines. {Right panels}: The simulations are shown in more detail
($30\times$ zoom of each panel) to show the timescale and structure of variations.
}
\label{fig:tts}
\end{center}
\end{figure}

To calculate predicted transit times, we numerically integrate our
baseline model, which consists of a central star of mass 
0.912~$M_\Sun$ surrounded by planets with periods and epochs given 
in Table~\ref{tab:p-params} (given at dynamical epoch BJD 2454170), and with masses of
8.7, 0.65, 16.1, 1.1 and 8.0 $M_\Earth$ (from least to greatest
orbital period), corresponding to the best-fit masses for \planetb\ and
\planetc, a guess of $M_p = M_{\Earth}\ (R_{p}/R_{\Earth})^{2.06}$ \citep{lissauer11b} 
for \planetd, and masses giving Earth's density for \koie\ and \koif. (We remind the
reader that we have not, in this paper, validated these two candidates as planets.  However, 
considering them as such for the purposes of evaluating dynamical stability is the conservative choice, since the presence
of 5 planets, as opposed to 3, is more likely to induce dynamical instabilities. \citet{fressin12} presents the validation of 
\koie\ and \koif, gives their sizes, from which the masses above are derived, and discusses constraints on their masses.)
The orbits are chosen to be initially circular, coplanar, and edge-on to
the line of sight.  The root mean square deviations of the model
transit times from the best-fit linear ephemeris projected over 8 years 
are approximately 3s, 76s, 9s, 95s and 10s (from least to
greatest orbital period), all significantly smaller than the
measurement precision shown in Figure~\ref{fig:tts}.

Next, we investigated long-term stability for this system by
integrating the baseline model with the hybrid algorithm in Mercury
\citep{chambers99}. As no close encounters were recorded, this algorithm
reduced to the symplectic algorithm \citep{wisdom91}, with
time steps of 0.1~days, roughly 2.7\% of the period of the innermost planet. 
Over the 10~Myr integration duration, there were no indications of
instability.  The orbital eccentricities fluctuated on the scale between
approximately $3\times10^{-5}$ (for \planetd) and $0.001$ (for \koie).  We
conclude that plausible, low-eccentricity models for the system are
stable over long timescales.

Finally, we performed an ensemble of N-body integrations using the
time-symmetric 4th order Hermite integrator \citep{kokubo98} implemented in
{\tt Swarm-NG}\footnote{www.astro.ufl.edu/$\sim$eford/code/swarm/} to estimate
the maximum plausible eccentricity for each planet consistent with
long-term stability. For each N-body
integration, we set four planets on circular orbits and assigned one
planet a non-zero eccentricity.  The eccentricity and pericenter
directions for the planet on a non-circular orbit were drawn from
uniform distributions.  The maximum for the uniform distribution of
eccentricities was chosen to be slightly larger than necessary for its
orbit to cross one of its neighbors.  We report $e_{\max}$, the
maximum initial eccentricity that resulted in a system with no close
encounters (within one mutual Hill radius) and semi-major axes (in a
Jacobi frame) which varied by less than 1\% for the duration of the
integrations.  Based on 100 integrations per planet and relatively
short integrations ($10^5$ years), we estimate $e_{\max}$ to be
0.19, 0.16, 0.16, 0.38 and 0.55 (from smallest to largest orbital
period).  Technically, we can not completely exclude larger
eccentricities, due to various assumptions (such as the planet masses,
coplanarity, prograde orbits, absence of false positives, and the potential for small
islands of stability at higher eccentricity).  Nevertheless, the
N-body integrations support the assumption of non-crossing orbits, as
the vast majority of systems with an eccentricity larger than
$e_{\max}$ are dynamical unstable.

\begin{deluxetable}{lcccc}
\tabletypesize{\scriptsize}
\tablecaption{Transit Times for \starname\ {\it We request that this table be published in electronic format.}}
\tablewidth{0pt}
\tablehead{
\colhead{ID} & \colhead{n} & \colhead{$t_n$}        & \colhead{TTV$_n$} & \colhead{$\sigma_n$} \\ 
\colhead{}    & \colhead{}      & \colhead{BJD-2454900} & \colhead{(d)}  & \colhead{(d)} 
   }
\startdata
  Kepler-20b & \multicolumn{4}{c}{$ 67.50027 + n \times 3.6961219$} \\ 
  Kepler-20b &   0 &     67.4942 &    -0.0061 &    0.0032 \\
  Kepler-20b &   1 &     71.1956 &    -0.0008 &    0.0035 \\
  Kepler-20b &   2 &     74.8946 &     0.0021 &    0.0041 \\
  Kepler-20b &   3 &     78.5836 &    -0.0050 &    0.0035 \\
  Kepler-20b &   4 &     82.2857 &     0.0009 &    0.0034 \\
  Kepler-20b &   5 &     85.9768 &    -0.0040 &    0.0032 \\
  Kepler-20b &   6 &     89.6783 &     0.0013 &    0.0045 \\
  Kepler-20b &   7 &     93.3689 &    -0.0042 &    0.0037 \\
  Kepler-20b &   8 &     97.0699 &     0.0007 &    0.0071 \\
  Kepler-20b &  10 &    104.4602 &    -0.0012 &    0.0039 \\
  Kepler-20b &  11 &    108.1582 &     0.0006 &    0.0028 \\
  Kepler-20b &  12 &    111.8565 &     0.0028 &    0.0049 \\
  Kepler-20b &  14 &    119.2447 &    -0.0013 &    0.0036 \\
  Kepler-20b &  15 &    122.9460 &     0.0039 &    0.0043 \\
  Kepler-20b &  16 &    126.6347 &    -0.0035 &    0.0044 \\
  Kepler-20b &  17 &    130.3393 &     0.0050 &    0.0034 \\
  Kepler-20b &  18 &    134.0237 &    -0.0068 &    0.0034 \\
  Kepler-20b &  19 &    137.7315 &     0.0049 &    0.0044 \\
  Kepler-20b &  20 &    141.4302 &     0.0075 &    0.0043 \\
  Kepler-20b &  21 &    145.1190 &     0.0001 &    0.0027 \\
  Kepler-20b &  22 &    148.8142 &    -0.0007 &    0.0034 \\
  Kepler-20b &  23 &    152.5058 &    -0.0052 &    0.0031 \\
  Kepler-20b &  24 &    156.2050 &    -0.0022 &    0.0034 \\
  Kepler-20b &  25 &    159.8994 &    -0.0039 &    0.0045 \\
  Kepler-20b &  27 &    167.2983 &     0.0027 &    0.0035 \\
  Kepler-20b &  28 &    170.9908 &    -0.0009 &    0.0036 \\
  Kepler-20b &  29 &    174.6887 &     0.0009 &    0.0078 \\
  Kepler-20b &  30 &    178.3859 &     0.0020 &    0.0028 \\
  Kepler-20b &  31 &    182.0814 &     0.0013 &    0.0038 \\
  Kepler-20b &  32 &    185.7734 &    -0.0027 &    0.0038 \\
  Kepler-20b &  33 &    189.4547 &    -0.0176 &    0.0103 \\
  Kepler-20b &  35 &    196.8662 &     0.0016 &    0.0038 \\
  Kepler-20b &  36 &    200.5665 &     0.0059 &    0.0036 \\
  Kepler-20b &  37 &    204.2595 &     0.0027 &    0.0026 \\
  Kepler-20b &  38 &    207.9564 &     0.0035 &    0.0032 \\
  Kepler-20b &  39 &    211.6447 &    -0.0043 &    0.0030 \\
  Kepler-20b &  40 &    215.3402 &    -0.0049 &    0.0054 \\
  Kepler-20b &  41 &    219.0422 &     0.0009 &    0.0040 \\
  Kepler-20b &  42 &    222.7391 &     0.0017 &    0.0048 \\
  Kepler-20b &  43 &    226.4399 &     0.0064 &    0.0052 \\
  Kepler-20b &  44 &    230.1275 &    -0.0022 &    0.0024 \\
  Kepler-20b &  45 &    233.8232 &    -0.0025 &    0.0031 \\
  Kepler-20b &  46 &    237.5259 &     0.0041 &    0.0043 \\
  Kepler-20b &  47 &    241.2178 &    -0.0002 &    0.0037 \\
  Kepler-20b &  48 &    244.9212 &     0.0071 &    0.0035 \\
  Kepler-20b &  49 &    248.6075 &    -0.0028 &    0.0033 \\
  Kepler-20b &  50 &    252.3057 &    -0.0007 &    0.0045 \\
  Kepler-20b &  52 &    259.6893 &    -0.0093 &    0.0043 \\
  Kepler-20b &  53 &    263.4017 &     0.0070 &    0.0046 \\
  Kepler-20b &  54 &    267.0944 &     0.0036 &    0.0039 \\
  Kepler-20b &  55 &    270.7832 &    -0.0038 &    0.0051 \\
  Kepler-20b &  56 &    274.4814 &    -0.0017 &    0.0042 \\
  Kepler-20b &  57 &    278.1816 &     0.0024 &    0.0049 \\
  Kepler-20b &  58 &    281.8728 &    -0.0026 &    0.0041 \\
  Kepler-20b &  59 &    285.5734 &     0.0020 &    0.0044 \\
  Kepler-20b &  60 &    289.2635 &    -0.0041 &    0.0041 \\
  Kepler-20b &  61 &    292.9654 &     0.0017 &    0.0053 \\
  Kepler-20b &  62 &    296.6559 &    -0.0040 &    0.0028 \\
  Kepler-20b &  63 &    300.3544 &    -0.0015 &    0.0039 \\
  Kepler-20b &  64 &    304.0540 &     0.0019 &    0.0035 \\
  Kepler-20b &  65 &    307.7479 &    -0.0002 &    0.0040 \\
  Kepler-20b &  66 &    311.4449 &     0.0005 &    0.0045 \\
  Kepler-20b &  67 &    315.1474 &     0.0070 &    0.0042 \\
  Kepler-20b &  68 &    318.8345 &    -0.0020 &    0.0045 \\
  Kepler-20b &  69 &    322.5374 &     0.0048 &    0.0040 \\
  Kepler-20b &  70 &    326.2274 &    -0.0014 &    0.0039 \\
  Kepler-20b &  74 &    341.0195 &     0.0062 &    0.0040 \\
  Kepler-20b &  75 &    344.7166 &     0.0072 &    0.0041 \\
  Kepler-20b &  76 &    348.4024 &    -0.0031 &    0.0048 \\
  Kepler-20b &  77 &    352.1038 &     0.0022 &    0.0038 \\
  Kepler-20b &  78 &    355.8025 &     0.0047 &    0.0029 \\
  Kepler-20b &  79 &    359.4904 &    -0.0035 &    0.0043 \\
  Kepler-20b &  80 &    363.1916 &     0.0016 &    0.0045 \\
  Kepler-20b &  81 &    366.8831 &    -0.0031 &    0.0041 \\
  Kepler-20b &  82 &    370.5829 &     0.0006 &    0.0037 \\
  Kepler-20b &  83 &    374.2740 &    -0.0044 &    0.0027 \\
  Kepler-20b &  84 &    377.9727 &    -0.0018 &    0.0040 \\
  Kepler-20b &  85 &    381.6793 &     0.0086 &    0.0039 \\
  Kepler-20b &  86 &    385.3678 &     0.0011 &    0.0033 \\
  Kepler-20b &  87 &    389.0654 &     0.0025 &    0.0041 \\
  Kepler-20b &  88 &    392.7595 &     0.0005 &    0.0045 \\
  Kepler-20b &  89 &    396.4618 &     0.0067 &    0.0042 \\
  Kepler-20b &  90 &    400.1460 &    -0.0053 &    0.0042 \\
  Kepler-20b &  91 &    403.8471 &    -0.0003 &    0.0043 \\
  Kepler-20b &  92 &    407.5457 &     0.0022 &    0.0027 \\
  Kepler-20b &  93 &    411.2340 &    -0.0056 &    0.0049 \\
  Kepler-20b &  94 &    414.9291 &    -0.0066 &    0.0036 \\
  Kepler-20b &  95 &    418.6325 &     0.0007 &    0.0034 \\
  Kepler-20b &  96 &    422.3282 &     0.0003 &    0.0029 \\
  Kepler-20b &  97 &    426.0243 &     0.0002 &    0.0017 \\
  Kepler-20b &  98 &    429.7166 &    -0.0036 &    0.0043 \\
  Kepler-20b &  99 &    433.4211 &     0.0047 &    0.0030 \\
  Kepler-20b & 101 &    440.8062 &    -0.0024 &    0.0049 \\
  Kepler-20b & 102 &    444.4994 &    -0.0053 &    0.0044 \\
  Kepler-20b & 103 &    448.2026 &     0.0018 &    0.0035 \\
  Kepler-20b & 104 &    451.9098 &     0.0129 &    0.0053 \\
  Kepler-20b & 105 &    455.5917 &    -0.0013 &    0.0045 \\
  Kepler-20b & 106 &    459.2855 &    -0.0037 &    0.0039 \\
  Kepler-20b & 107 &    462.9828 &    -0.0025 &    0.0027 \\
  Kepler-20b & 108 &    466.6899 &     0.0084 &    0.0038 \\
  Kepler-20b & 109 &    470.3763 &    -0.0012 &    0.0040 \\
  Kepler-20b & 110 &    474.0741 &     0.0004 &    0.0041 \\
  Kepler-20b & 111 &    477.7694 &    -0.0004 &    0.0033 \\
  Kepler-20b & 112 &    481.4629 &    -0.0030 &    0.0034 \\
  Kepler-20b & 113 &    485.1567 &    -0.0053 &    0.0034 \\
  Kepler-20b & 114 &    488.8559 &    -0.0023 &    0.0025 \\
  Kepler-20b & 115 &    492.5534 &    -0.0009 &    0.0035 \\
  Kepler-20b & 116 &    496.2526 &     0.0022 &    0.0033 \\
  Kepler-20b & 118 &    503.6411 &    -0.0015 &    0.0039 \\
  Kepler-20b & 119 &    507.3215 &    -0.0173 &    0.0045 \\
  Kepler-20b & 120 &    511.0371 &     0.0022 &    0.0055 \\
  Kepler-20b & 121 &    514.7318 &     0.0008 &    0.0049 \\
  Kepler-20b & 122 &    518.4255 &    -0.0016 &    0.0027 \\
  Kepler-20b & 123 &    522.1195 &    -0.0037 &    0.0043 \\
  Kepler-20b & 124 &    525.8160 &    -0.0034 &    0.0030 \\
  Kepler-20b & 125 &    529.5154 &    -0.0001 &    0.0044 \\
  Kepler-20b & 126 &    533.2147 &     0.0031 &    0.0046 \\
  Kepler-20b & 127 &    536.9038 &    -0.0039 &    0.0029 \\
  Kepler-20b & 128 &    540.6034 &    -0.0004 &    0.0037 \\
  Kepler-20b & 129 &    544.3084 &     0.0084 &    0.0042 \\
  Kepler-20b & 130 &    548.0018 &     0.0057 &    0.0041 \\
  Kepler-20b & 131 &    551.6946 &     0.0024 &    0.0035 \\
  Kepler-20b & 132 &    555.3912 &     0.0028 &    0.0030 \\
  Kepler-20b & 133 &    559.0861 &     0.0016 &    0.0037 \\
  Kepler-20b & 135 &    566.4817 &     0.0050 &    0.0043 \\
  Kepler-20b & 136 &    570.1738 &     0.0009 &    0.0054 \\
  Kepler-20b & 137 &    573.8736 &     0.0046 &    0.0032 \\
  Kepler-20b & 138 &    577.5612 &    -0.0039 &    0.0040 \\
  Kepler-20b & 139 &    581.2539 &    -0.0074 &    0.0051 \\
  Kepler-20b & 140 &    584.9482 &    -0.0092 &    0.0040 \\
  Kepler-20b & 141 &    588.6502 &    -0.0032 &    0.0035 \\
  Kepler-20b & 142 &    592.3510 &     0.0014 &    0.0050 \\
  Kepler-20b & 143 &    596.0490 &     0.0033 &    0.0034 \\
  Kepler-20b & 144 &    599.7493 &     0.0075 &    0.0044 \\
  Kepler-20b & 145 &    603.4392 &     0.0013 &    0.0056 \\
  Kepler-20b & 146 &    607.1344 &     0.0004 &    0.0038 \\
  Kepler-20b & 147 &    610.8329 &     0.0027 &    0.0043 \\
  Kepler-20b & 148 &    614.5286 &     0.0023 &    0.0032 \\
  Kepler-20b & 149 &    618.2205 &    -0.0019 &    0.0027 \\
  Kepler-20b & 150 &    621.9156 &    -0.0029 &    0.0031 \\
  Kepler-20b & 151 &    625.6180 &     0.0033 &    0.0039 \\
  Kepler-20b & 152 &    629.3115 &     0.0007 &    0.0039 \\
  Kepler-20b & 153 &    633.0118 &     0.0049 &    0.0050 \\
  Kepler-20b & 154 &    636.6958 &    -0.0072 &    0.0043 \\
  Kepler-20b & 155 &    640.3937 &    -0.0054 &    0.0045 \\
  Kepler-20b & 157 &    647.7922 &     0.0008 &    0.0046 \\
  Kepler-20b & 158 &    651.4935 &     0.0060 &    0.0042 \\
  Kepler-20b & 163 &    669.9606 &    -0.0075 &    0.0041 \\
  Kepler-20b & 164 &    673.6626 &    -0.0017 &    0.0094 \\
  Kepler-20b & 165 &    677.3552 &    -0.0052 &    0.0037 \\
  Kepler-20b & 166 &    681.0550 &    -0.0015 &    0.0062 \\
  Kepler-20b & 167 &    684.7541 &     0.0014 &    0.0030 \\
  Kepler-20b & 168 &    688.4526 &     0.0038 &    0.0046 \\
  Kepler-20b & 169 &    692.1463 &     0.0014 &    0.0017 \\
  Kepler-20b & 171 &    699.5324 &    -0.0047 &    0.0039 \\
  Kepler-20b & 172 &    703.2330 &    -0.0002 &    0.0067 \\
  Kepler-20b & 173 &    706.9307 &     0.0014 &    0.0038 \\
  Kepler-20b & 174 &    710.6210 &    -0.0045 &    0.0032 \\
  Kepler-20b & 175 &    714.3175 &    -0.0041 &    0.0037 \\
  Kepler-20b & 176 &    718.0073 &    -0.0105 &    0.0063 \\
  Kepler-20b & 177 &    721.7094 &    -0.0044 &    0.0042 \\
  Kepler-20b & 178 &    725.4097 &    -0.0003 &    0.0039 \\
  Kepler-20b & 179 &    729.1068 &     0.0007 &    0.0039 \\
  Kepler-20b & 180 &    732.8057 &     0.0035 &    0.0032 \\

  70.04 & \multicolumn{4}{c}{$ 68.9336 + n \times 6.098493$} \\ 
  70.04 &   0 &     68.9295 &    -0.0041 &    0.0124 \\
  70.04 &   1 &     75.0316 &    -0.0005 &    0.0149 \\
  70.04 &   2 &     81.1487 &     0.0181 &    0.0116 \\
  70.04 &   3 &     87.1926 &    -0.0364 &    0.0176 \\
  70.04 &   4 &     93.2997 &    -0.0278 &    0.0117 \\
  70.04 &   6 &    105.5245 &    -0.0001 &    0.0103 \\
  70.04 &   7 &    111.6445 &     0.0215 &    0.0150 \\
  70.04 &   8 &    117.6964 &    -0.0251 &    0.0150 \\
  70.04 &   9 &    123.8323 &     0.0123 &    0.0138 \\
  70.04 &  10 &    129.9095 &    -0.0090 &    0.0122 \\
  70.04 &  11 &    135.9953 &    -0.0217 &    0.0525 \\
  70.04 &  12 &    142.0980 &    -0.0175 &    0.0176 \\
  70.04 &  13 &    148.2067 &    -0.0073 &    0.0211 \\
  70.04 &  14 &    154.3205 &     0.0080 &    0.0156 \\
  70.04 &  15 &    160.4350 &     0.0240 &    0.0122 \\
  70.04 &  16 &    166.5186 &     0.0091 &    0.0152 \\
  70.04 &  17 &    172.6143 &     0.0063 &    0.0123 \\
  70.04 &  18 &    178.7111 &     0.0047 &    0.0141 \\
  70.04 &  19 &    184.8108 &     0.0059 &    0.0503 \\
  70.04 &  20 &    190.9342 &     0.0307 &    0.0171 \\
  70.04 &  21 &    196.9966 &    -0.0054 &    0.0137 \\
  70.04 &  22 &    203.1011 &     0.0007 &    0.0148 \\
  70.04 &  23 &    209.2095 &     0.0106 &    0.0177 \\
  70.04 &  24 &    215.3090 &     0.0115 &    0.0181 \\
  70.04 &  25 &    221.3969 &     0.0010 &    0.0163 \\
  70.04 &  26 &    227.5457 &     0.0513 &    0.0112 \\
  70.04 &  27 &    233.6113 &     0.0183 &    0.0132 \\
  70.04 &  28 &    239.7084 &     0.0170 &    0.0226 \\
  70.04 &  29 &    245.7739 &    -0.0160 &    0.0192 \\
  70.04 &  30 &    251.8802 &    -0.0082 &    0.0147 \\
  70.04 &  31 &    257.9671 &    -0.0198 &    0.0147 \\
  70.04 &  32 &    264.0366 &    -0.0488 &    0.0175 \\
  70.04 &  33 &    270.1797 &    -0.0042 &    0.0314 \\
  70.04 &  34 &    276.2872 &     0.0048 &    0.0161 \\
  70.04 &  35 &    282.3650 &    -0.0159 &    0.0197 \\
  70.04 &  36 &    288.4550 &    -0.0244 &    0.0211 \\
  70.04 &  37 &    294.5290 &    -0.0489 &    0.0186 \\
  70.04 &  38 &    300.6801 &     0.0037 &    0.0163 \\
  70.04 &  39 &    306.7537 &    -0.0211 &    0.0169 \\
  70.04 &  40 &    312.9025 &     0.0292 &    0.0190 \\
  70.04 &  41 &    318.9743 &     0.0024 &    0.0131 \\
  70.04 &  42 &    325.0517 &    -0.0186 &    0.0129 \\
  70.04 &  45 &    343.4285 &     0.0627 &    0.0183 \\
  70.04 &  46 &    349.4757 &     0.0114 &    0.0189 \\
  70.04 &  47 &    355.5391 &    -0.0236 &    0.0120 \\
  70.04 &  48 &    361.6421 &    -0.0192 &    0.0113 \\
  70.04 &  49 &    367.7477 &    -0.0120 &    0.0104 \\
  70.04 &  50 &    373.8366 &    -0.0217 &    0.0106 \\
  70.04 &  51 &    379.9503 &    -0.0065 &    0.0109 \\
  70.04 &  52 &    386.0611 &     0.0059 &    0.0128 \\
  70.04 &  53 &    392.1451 &    -0.0086 &    0.0139 \\
  70.04 &  54 &    398.2748 &     0.0226 &    0.0204 \\
  70.04 &  55 &    404.3426 &    -0.0082 &    0.0139 \\
  70.04 &  57 &    416.5261 &    -0.0216 &    0.0164 \\
  70.04 &  58 &    422.6521 &     0.0059 &    0.0142 \\
  70.04 &  59 &    428.7752 &     0.0305 &    0.0155 \\
  70.04 &  60 &    434.8565 &     0.0133 &    0.0164 \\
  70.04 &  61 &    440.9316 &    -0.0101 &    0.0151 \\
  70.04 &  62 &    447.0528 &     0.0127 &    0.0172 \\
  70.04 &  63 &    453.1351 &    -0.0035 &    0.0135 \\
  70.04 &  64 &    459.2399 &     0.0028 &    0.0119 \\
  70.04 &  65 &    465.3240 &    -0.0116 &    0.0147 \\
  70.04 &  67 &    477.5344 &     0.0017 &    0.0120 \\
  70.04 &  68 &    483.6637 &     0.0326 &    0.0097 \\
  70.04 &  69 &    489.7101 &    -0.0195 &    0.0152 \\
  70.04 &  70 &    495.8219 &    -0.0062 &    0.0158 \\
  70.04 &  71 &    501.9295 &     0.0029 &    0.0135 \\
  70.04 &  72 &    507.9607 &    -0.0644 &    0.0155 \\
  70.04 &  73 &    514.1161 &    -0.0075 &    0.0148 \\
  70.04 &  74 &    520.1992 &    -0.0229 &    0.0127 \\
  70.04 &  75 &    526.3201 &    -0.0005 &    0.0121 \\
  70.04 &  76 &    532.4389 &     0.0198 &    0.0133 \\
  70.04 &  77 &    538.5127 &    -0.0049 &    0.0143 \\
  70.04 &  78 &    544.6282 &     0.0121 &    0.0113 \\
  70.04 &  79 &    550.7134 &    -0.0011 &    0.0182 \\
  70.04 &  80 &    556.8084 &    -0.0046 &    0.0092 \\
  70.04 &  82 &    568.9873 &    -0.0227 &    0.0189 \\
  70.04 &  83 &    575.1346 &     0.0261 &    0.0154 \\
  70.04 &  84 &    581.2328 &     0.0258 &    0.0157 \\
  70.04 &  85 &    587.2972 &    -0.0083 &    0.0147 \\
  70.04 &  87 &    599.4992 &    -0.0033 &    0.0139 \\
  70.04 &  88 &    605.6078 &     0.0068 &    0.0158 \\
  70.04 &  89 &    611.6560 &    -0.0435 &    0.0111 \\
  70.04 &  90 &    617.8120 &     0.0141 &    0.0169 \\
  70.04 &  92 &    630.0025 &     0.0075 &    0.0151 \\
  70.04 &  93 &    636.1181 &     0.0247 &    0.0147 \\
  70.04 &  94 &    642.1929 &     0.0010 &    0.0106 \\
  70.04 &  95 &    648.3022 &     0.0118 &    0.0171 \\
  70.04 &  99 &    672.6873 &     0.0029 &    0.0192 \\
  70.04 & 100 &    678.7764 &    -0.0065 &    0.0153 \\
  70.04 & 101 &    684.8819 &     0.0006 &    0.0144 \\
  70.04 & 102 &    690.9771 &    -0.0027 &    0.0165 \\
  70.04 & 103 &    697.0837 &     0.0053 &    0.0166 \\
  70.04 & 104 &    703.1848 &     0.0079 &    0.0228 \\
  70.04 & 105 &    709.3466 &     0.0713 &    0.0171 \\
  70.04 & 106 &    715.3852 &     0.0113 &    0.0108 \\
  70.04 & 107 &    721.4592 &    -0.0132 &    0.0155 \\
  70.04 & 108 &    727.5689 &    -0.0019 &    0.0169 \\
  70.04 & 109 &    733.6590 &    -0.0104 &    0.0103 \\

  Kepler-20c & \multicolumn{4}{c}{$ 71.6076 + n \times 10.85409$} \\ 
  Kepler-20c &   0 &     71.6063 &    -0.0013 &    0.0018 \\
  Kepler-20c &   1 &     82.4610 &    -0.0007 &    0.0027 \\
  Kepler-20c &   2 &     93.3240 &     0.0083 &    0.0033 \\
  Kepler-20c &   3 &    104.1716 &     0.0017 &    0.0024 \\
  Kepler-20c &   5 &    125.8775 &    -0.0005 &    0.0020 \\
  Kepler-20c &   6 &    136.7324 &     0.0003 &    0.0018 \\
  Kepler-20c &   7 &    147.5848 &    -0.0014 &    0.0018 \\
  Kepler-20c &   8 &    158.4418 &     0.0015 &    0.0013 \\
  Kepler-20c &   9 &    169.2901 &    -0.0043 &    0.0014 \\
  Kepler-20c &  10 &    180.1483 &    -0.0002 &    0.0018 \\
  Kepler-20c &  11 &    191.0000 &    -0.0025 &    0.0018 \\
  Kepler-20c &  12 &    201.8560 &    -0.0006 &    0.0015 \\
  Kepler-20c &  13 &    212.7127 &     0.0020 &    0.0016 \\
  Kepler-20c &  14 &    223.5737 &     0.0089 &    0.0034 \\
  Kepler-20c &  15 &    234.4203 &     0.0014 &    0.0023 \\
  Kepler-20c &  16 &    245.2744 &     0.0013 &    0.0013 \\
  Kepler-20c &  18 &    266.9898 &     0.0086 &    0.0027 \\
  Kepler-20c &  19 &    277.8388 &     0.0034 &    0.0018 \\
  Kepler-20c &  20 &    288.6911 &     0.0017 &    0.0019 \\
  Kepler-20c &  21 &    299.5445 &     0.0009 &    0.0018 \\
  Kepler-20c &  22 &    310.3963 &    -0.0013 &    0.0016 \\
  Kepler-20c &  23 &    321.2501 &    -0.0016 &    0.0018 \\
  Kepler-20c &  25 &    342.9608 &     0.0009 &    0.0018 \\
  Kepler-20c &  26 &    353.8129 &    -0.0011 &    0.0014 \\
  Kepler-20c &  27 &    364.6678 &    -0.0003 &    0.0014 \\
  Kepler-20c &  29 &    386.3751 &    -0.0011 &    0.0020 \\
  Kepler-20c &  30 &    397.2301 &    -0.0002 &    0.0013 \\
  Kepler-20c &  32 &    418.9392 &     0.0007 &    0.0021 \\
  Kepler-20c &  33 &    429.7823 &    -0.0103 &    0.0032 \\
  Kepler-20c &  34 &    440.6474 &     0.0007 &    0.0032 \\
  Kepler-20c &  35 &    451.5008 &     0.0000 &    0.0012 \\
  Kepler-20c &  36 &    462.3549 &     0.0000 &    0.0022 \\
  Kepler-20c &  37 &    473.2071 &    -0.0019 &    0.0015 \\
  Kepler-20c &  38 &    484.0631 &     0.0000 &    0.0017 \\
  Kepler-20c &  39 &    494.9177 &     0.0006 &    0.0029 \\
  Kepler-20c &  40 &    505.7737 &     0.0024 &    0.0011 \\
  Kepler-20c &  41 &    516.6244 &    -0.0010 &    0.0016 \\
  Kepler-20c &  42 &    527.4780 &    -0.0015 &    0.0013 \\
  Kepler-20c &  43 &    538.3328 &    -0.0008 &    0.0014 \\
  Kepler-20c &  44 &    549.1875 &    -0.0001 &    0.0021 \\
  Kepler-20c &  45 &    560.0406 &    -0.0011 &    0.0015 \\
  Kepler-20c &  46 &    570.8982 &     0.0024 &    0.0014 \\
  Kepler-20c &  47 &    581.7495 &    -0.0004 &    0.0021 \\
  Kepler-20c &  48 &    592.6034 &    -0.0006 &    0.0034 \\
  Kepler-20c &  49 &    603.4558 &    -0.0023 &    0.0033 \\
  Kepler-20c &  50 &    614.3101 &    -0.0021 &    0.0023 \\
  Kepler-20c &  51 &    625.1658 &    -0.0005 &    0.0014 \\
  Kepler-20c &  52 &    636.0218 &     0.0014 &    0.0014 \\
  Kepler-20c &  53 &    646.8745 &     0.0000 &    0.0016 \\
  Kepler-20c &  55 &    668.5835 &     0.0009 &    0.0016 \\
  Kepler-20c &  56 &    679.4363 &    -0.0005 &    0.0025 \\
  Kepler-20c &  57 &    690.2898 &    -0.0010 &    0.0023 \\
  Kepler-20c &  58 &    701.1466 &     0.0017 &    0.0023 \\
  Kepler-20c &  59 &    711.9975 &    -0.0015 &    0.0018 \\
  Kepler-20c &  60 &    722.8602 &     0.0071 &    0.0024 \\
  Kepler-20c &  61 &    733.7038 &    -0.0034 &    0.0014 \\

  70.05 & \multicolumn{4}{c}{$ 68.219 + n \times 19.57706$} \\ 
  70.05 &   0 &     68.2114 &    -0.0076 &    0.0111 \\
  70.05 &   1 &     87.7637 &    -0.0323 &    0.0155 \\
  70.05 &   2 &    107.3404 &    -0.0327 &    0.0149 \\
  70.05 &   3 &    126.9232 &    -0.0270 &    0.0161 \\
  70.05 &   4 &    146.5711 &     0.0439 &    0.0177 \\
  70.05 &   5 &    166.0304 &    -0.0739 &    0.0121 \\
  70.05 &   6 &    185.6525 &    -0.0289 &    0.0138 \\
  70.05 &   7 &    205.2436 &    -0.0149 &    0.0191 \\
  70.05 &   8 &    224.8237 &    -0.0117 &    0.0134 \\
  70.05 &   9 &    244.4134 &     0.0009 &    0.0259 \\
  70.05 &  10 &    264.0047 &     0.0151 &    0.0129 \\
  70.05 &  12 &    303.1682 &     0.0245 &    0.0134 \\
  70.05 &  13 &    322.7048 &    -0.0160 &    0.0135 \\
  70.05 &  14 &    342.2962 &    -0.0017 &    0.0119 \\
  70.05 &  15 &    361.8658 &    -0.0091 &    0.0217 \\
  70.05 &  16 &    381.4339 &    -0.0181 &    0.0154 \\
  70.05 &  17 &    401.0316 &     0.0025 &    0.0157 \\
  70.05 &  18 &    420.5948 &    -0.0113 &    0.0129 \\
  70.05 &  19 &    440.1844 &     0.0013 &    0.0128 \\
  70.05 &  20 &    459.7616 &     0.0014 &    0.0193 \\
  70.05 &  21 &    479.3348 &    -0.0024 &    0.0110 \\
  70.05 &  22 &    498.9270 &     0.0127 &    0.0138 \\
  70.05 &  23 &    518.4884 &    -0.0030 &    0.0113 \\
  70.05 &  24 &    538.0258 &    -0.0427 &    0.0118 \\
  70.05 &  25 &    557.6534 &     0.0079 &    0.0114 \\
  70.05 &  26 &    577.2165 &    -0.0061 &    0.0207 \\
  70.05 &  27 &    596.8058 &     0.0062 &    0.0157 \\
  70.05 &  28 &    616.4009 &     0.0242 &    0.0148 \\
  70.05 &  29 &    635.9785 &     0.0247 &    0.0145 \\
  70.05 &  31 &    675.1005 &    -0.0074 &    0.0192 \\
  70.05 &  33 &    714.2609 &    -0.0010 &    0.0139 \\
  70.05 &  34 &    733.8455 &     0.0065 &    0.0101 \\

  Kepler-20d & \multicolumn{4}{c}{$ 97.7271 + n \times 77.61184$} \\ 
  Kepler-20d &   0 &     97.7293 &     0.0022 &    0.0029 \\
  Kepler-20d &   1 &    175.3387 &    -0.0002 &    0.0022 \\
  Kepler-20d &   2 &    252.9482 &    -0.0026 &    0.0040 \\
  Kepler-20d &   5 &    485.7834 &    -0.0029 &    0.0022 \\
  Kepler-20d &   6 &    563.4005 &     0.0024 &    0.0029 \\
  Kepler-20d &   7 &    641.0100 &     0.0001 &    0.0022 \\
  Kepler-20d &   8 &    718.6231 &     0.0013 &    0.0026 \\

\enddata
\label{tab:ttv}
\end{deluxetable}

\clearpage

\section{Constraints on the Planetary Compositions and Formation History}

\begin{figure}
\begin{center}
\includegraphics[scale=0.65]{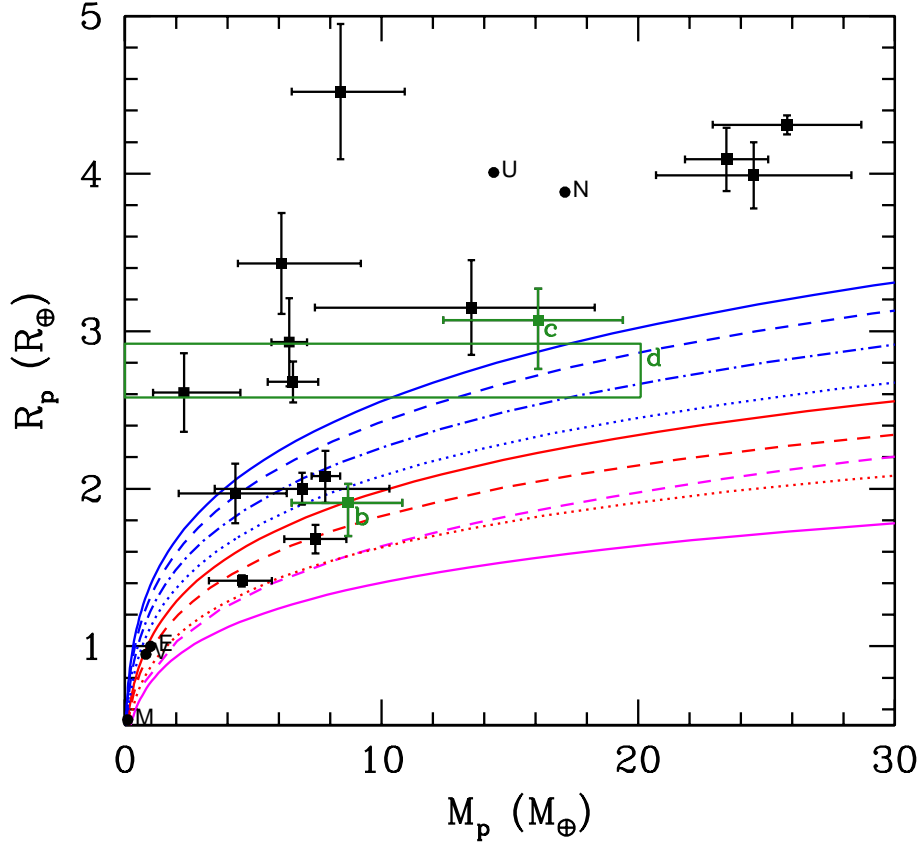}
\caption{Mass-radius relationships of small transiting planets. The
  three confirmed planets in the \starname\ system are highlighted in
  green. \planetb\ and \planetc\ are plotted with error bars
  delimiting the $1~\sigma$ uncertainties on the planet mass and
  radius, while \planetd\ is plotted with bands illustrating the $2~\sigma$ mass upper limit. Other small transiting exoplanets with measured masses (Kepler-10b, CoRoT-7b, Kepler-11bcdef, Kepler-18b, 55Cnc e, GJ~1214b, HD~97658b, GJ~436b, Kepler-4b, HAT-P-11b) are plotted in black. The Solar System planets are indicated with the first letters of their names.
The curves are illustrative constant-temperature (300~K) mass-radius
relations for bodies devoid of H/He from \citet{seager07}. The solid
lines are homogeneous-composition planets: water ice (blue solid),
MgSiO$_3$ perovskite (red solid), and iron (magenta solid). The
non-solid lines are mass-radius relations for differentiated planets:
75\% water ice, 22\% silicate shell, and 3\% iron core (blue dashed);
Ganymede-like with 45\% water ice, 48.5\% silicate shell, and 6.5\%
iron core (blue dot-dashed); 25\% water ice, 52.5\% silicate shell,
and 22.5\% iron core (blue dotted); Earth-like with 67.5\% silicate
mantle and 32.5\% iron core (red dashed); and Mercury-like with 30\%
silicate mantle and 70\% iron core (red dotted). The minimal radius
curve based on simulations of collisional mantle stripping from
differentiated silicate-iron planets \citep{marcus10} is denoted by
the dashed magenta line.}
\label{fig:rogersmr}
\end{center}
\end{figure}

The \starname\ system, harboring multiple sub-Neptune planets with
constrained radii and masses, informs our understanding of both models of
planet formation and the interior structure of planets that straddle
the boundary between sub-Neptunes and super-Earths. 
The transit radii measured by {\it Kepler} and the planetary masses
measured (or bounded) by radial velocity observations together
constrain the interior compositions of \planetb, \planetc, and
\planetd, as illustrated by the mass-radius diagram
(Figure~\ref{fig:rogersmr}).  We employ planet interior structure models
\citep{rogers10, rogers11} to explore the range of plausible planet
compositions. The interpretation is challenging because we do not yet
know if these sub-Neptune planets had a stunted formation, or if they formed as gas giants and then lost significant mass to evaporation \citep{baraffe04}. This is partly owing to the uncertainties involved in atmospheric escape modeling.

Notably, both \planetc\ and \planetd\ require significant volatile contents to account for their low mean densities, and cannot be composed of rocky and iron material alone. 
The volatile material in these planets could take the form of ices (H$_2$O, CH$_4$, NH$_3$) and/or H/He gas accreted during planet formation. Outgassing of rocky planets releases an insufficient quantity of volatiles (no more than 23\% H$_2$O and 3.6\% H$_2$ relative to the planet mass) to account for \planetc\ and could account for \planetd\ only in fine-tuned near-optimal outgassing scenarios \citep{elkinstanton08, schaefer08, rogers11}.
 For \planetc, ices (likely dominated by H$_2$O)  would need to constitute the majority of its mass, in the absence of a voluminous, though low-mass, envelope of light gases. Alternatively, a composition with approximately 1\% by mass H/He surrounding an Earth-composition refractory interior also matches the observed properties of the planet within $1~\sigma$. Intermediate scenarios, wherein both H/He and higher mean molecular weight volatile species from ices contribute to the planet mass, are also possible. 
For \planetd, the $2~\sigma$ upper limit on the planet density demands
at least a few percent H$_2$O by mass, or a few tenths of a percent H/He by mass. 

The nature of \planetb's composition is ambiguous: \planetb\  could be
terrestrial (with the transit radius defined by a rocky surface), or
it could support a significant gas envelope (like \planetc\ and
\planetd). In the mass-radius diagram (Figure~\ref{fig:rogersmr}), the
measured properties of \planetb\ straddle the pure-silicate
composition curve that defines a strict upper bound to rocky planet
radii. If \planetb\ is in fact a terrestrial planet consisting of an
iron core surrounded by a silicate mantle, the $1~\sigma$ limits on
the planet mass and radius constrain the iron core to be less than
62\% of the planet mass. In particular, an Earth-like composition
(30\% iron core, 70\% silicate mantle) is possible and matches the
observational constraints to within 1~$\sigma$, but a Mercury-like
composition (70\% iron core, 30\% silicate mantle) is not acceptable. Alternatively, \planetb\ may harbor a substantial gas layer like its sibling planets \planetc\ and \planetd\ at  larger orbital semi-major axes, and/or contain a significant 
component of astrophysical ices such as H$_{2}$O. The $1~\sigma$ lower limits on the planet density constrain the fraction of \planetb's mass that can be contributed by H$_2$O ($\lesssim55\%$) and H/He ($\lesssim1\%$).

Given their high levels of stellar irradiation (the semi-major axes of all five \starname\ planets are smaller than that of Mercury), atmospheric escape likely played an important role sculpting the compositions of the \starname\ planets. Planet compositions with low mean molecular weight gas envelopes would be especially susceptible to mass loss. 
Using a model for energy limited escape from hydrogen-rich envelopes \citep{lecavelier07}, we estimate that \planetb\ would be losing on the order of $4\times10^6~{\rm kg\, s^{-1}}$, which corresponds to 0.02~$M_{\oplus} {\rm Gyr^{-1}}$. Following the same approach, the estimated hydrogen mass loss rates for \planetc\ and \planetd\ are $2\times10^6~{\rm kg\, s^{-1}} $ ($0.01~{ M_{\oplus} {\rm Gyr^{-1}}}$) and $8\times10^4~{\rm kg\, s^{-1}}$ ($0.0004~{ M_{\oplus} {\rm Gyr^{-1}}}$), respectively. Our theoretical understanding of atmospheric escape from highly irradiated super-Earth and sub-Neptune exoplanets is very uncertain, and higher mass loss rates are plausible (especially at earlier times when the host star was more active). It is intriguing that \planetb, with its shorter orbital period and greater vulnerability to mass loss, also has a higher mean density than \planetc\ and \planetd.  More detailed modeling may constrain \planetb's compositional history and the extent to which its relative paucity of volatiles can be attributed to evaporation.

The  \starname\ planetary system shares several remarkable attributes with Kepler-11 \citep{lissauer11b}, namely the presence of multiple transiting low-density low-mass planets in a closely spaced orbital architecture.  The \starname\ system is less extreme than Kepler-11 in the realms of both low planet densities (Figure~\ref{fig:rogersmr}) and dynamical compactness (the Kepler-11 planets exhibit TTVs while the \starname\ planets do not). 

A striking feature of the \starname\ planetary system is the presence of Earth-size rocky planet candidates interspersed between volatile-rich sub-Neptunes at smaller and larger orbital semi-major axes, as also seen in \kepler\ candidate multi-planet systems \citep{lissauer11a}. 
Assuming that both \koie\ and \koif\ are planets, the distribution of the \starname\ planets in orbital order is as follows: \planetb\ (3.7~days, 1.9~$R_{\oplus}$), \koie\ (6.1~days, 0.9~$R_{\oplus}$), \planetc\ (10.9~days, 3.1~$R_{\oplus}$), \koif\ (19.6~days, 1.0~$R_{\oplus}$), and \planetd\ (77.6~days, 2.8~$R_{\oplus}$).  Given the radii and irradiation fluxes of the two Earth-size planet candidates, they would not retain gas envelopes. The first, second, and fourth planets have high densities indicative of solid planets, while the other two planets have low densities requiring significant volatile content. The volatile-rich third planet, \planetc\, dominates the inner part of the Kepler-20 system, by holding much more mass than the other three inner planets put together. In the Solar System, the terrestrial planets, gas-giants, and ice giants are neatly segregated in regions with increasing distance from the sun. Planet formation theories were developed to retrodict these Solar System composition trends \citep[e.g.,][]{safronov69, chambers10, dangelo10}. In the \starname\ system, the locations of the low-density sub-Neptunes that are rich in water and/or gas, and the Earth-size planet candidates 
does not exhibit a clean ordering with orbital period, challenging the conventional planet formation paradigm. In situ assembly may form multi-planet systems with close-in 
hot-Neptunes and super-Earths, provided the initial protoplanetary disk contained massive  amounts of solids ($\sim 50$--100~$M_{\oplus}$) within 1AU of the star \citep{hansen11}.

\acknowledgments
{\it Kepler} was competitively selected as the tenth Discovery mission. Funding for this mission is provided by NASA's Science Mission Directorate. The authors thank many people who gave so generously of their time to make this mission a success. This work is also based in part on observations made with the {\it Spitzer Space Telescope}, which is operated by the Jet Propulsion Laboratory, California Institute of Technology under a contract with NASA. Support for this work was provided by NASA through an award issued by JPL/Caltech. We would like to thank the {\it Spitzer} staff at IPAC and in particular Nancy Silbermann for scheduling the Spitzer observations of this program. Some of the data presented herein were obtained at the W.~M. Keck Observatory, which is operated as a scientific partnership among the California Institute of Technology, the University of California and the National Aeronautics and Space Administration. The Observatory was made possible by the generous financial support of the W.~M. Keck Foundation. (c) 2011 all rights reserved.

\end{document}